\documentclass[a4paper,11pt]{article}
\pdfoutput=1 

\usepackage{jcappub} 

\usepackage[T1]{fontenc} 
\usepackage{hyperref}

\def\pfrac#1#2{\left( \frac{#1}{#2} \right)}
\def\iso#1#2{\mbox{${}^{#2}{\rm #1}$}}
\def\he#1{\iso{He}{#1}}
\def\li#1{\iso{Li}{#1}}
\def\be#1{\iso{Be}{#1}}
\def\beq{\begin{equation}}
\def\eeq{\end{equation}}
\def\beqar{\begin{eqnarray}}
\def\eeqar{\end{eqnarray}}

\begin{document}

\begin{flushright}
UMN--TH--4308/24, FTPI--MINN--24/01   \\
January 2024
\end{flushright}

\title{Limits on Non-Relativistic Matter During Big-Bang Nucleosynthesis
}


\author[a,1]{Tsung-Han Yeh,\note{Corresponding author.}}
\author[b]{Keith A. Olive}
\author[c,d,e]{Brian D. Fields}


\affiliation[a]{TRIUMF, 4004 Wesbrook Mall, Vancouver, BC V6T 2A3, Canada}
\affiliation[b]{William I. Fine Theoretical Physics Institute, School of
 Physics and Astronomy, University of Minnesota, Minneapolis, MN 55455}
\affiliation[c]{Department of Astronomy, University of Illinois, Urbana, IL 61801}
\affiliation[d]{Department of Physics, University of Illinois, Urbana IL 61801}
\affiliation[e]{Illinois Center for Advanced Studies of the Universe}

\emailAdd{thyeh@triumf.ca}

\abstract{Big-bang nucleosynthesis (BBN) probes the cosmic mass-energy density at temperatures $\sim 10$ MeV to $\sim 100$ keV.  Here, we consider the effect of a cosmic matter-like species that is non-relativistic and pressureless during BBN.  Such a component must decay; doing so during BBN can alter the baryon-to-photon ratio, $\eta$, and the effective number of neutrino species.  We use light element abundances and the cosmic microwave background (CMB) constraints on $\eta$ and $N_\nu$ to place constraints on such a matter component.
We find that electromagnetic decays heat the photons relative to neutrinos, and thus dilute the effective number of relativistic species
to $N_{\rm eff} < 3$ for
the case of three Standard Model neutrino species.   Intriguingly,
likelihood results based on {\em Planck} CMB data alone find
$N_{\nu} = 2.800 \pm 0.294$, and when combined with standard BBN and the observations of D and \he4 give $N_{\nu} = 2.898 \pm 0.141$.
While both results are consistent with the Standard Model, we find that a nonzero abundance of electromagnetically decaying matter gives a better fit to these results.
Our best-fit results are for a matter species that decays entirely electromagnetically with a lifetime $\tau_X = 0.89 \ \rm sec$ and pre-decay density that is a fraction
$\xi = (\rho_X/\rho_{\rm rad})|_{10 \ \rm MeV} = 0.0026$ of the radiation energy density at 10 MeV; similarly good fits are found
over a range where $\xi \tau_X^{1/2}$ is constant.
On the other hand, decaying matter often spoils the BBN+CMB concordance, and we present limits in the $(\tau_X,\xi)$ plane
for both electromagnetic and invisible decays.  For dark (invisible) decays, standard BBN (i.e. $\xi=0$)
supplies the best fit. 
We end with a brief discussion of the impact of future measurements including CMB-S4. }

\maketitle
\flushbottom

\section{Introduction}
\label{sec:intro}

The successful concordance between big bang nucleosynthesis (BBN) \cite{bbn,osw,gary,iocco,CFOY,coc18,foyy,dpg,nlife}, the observational determination of the light elements D/H \cite{pc,cooke,riemer,bala,cookeN,riemer17,zava,CPS}, and \he4 \cite{aos4,abopss,Aver:2021rwi,Hsyu:2020uqb,Kurichin:2021ppm,vpp}, and observations of the cosmic microwave background (CMB) anisotropies \cite{wmap,Planck2018} is one of the foundations of early universe big bang cosmology. BBN is sensitive to all four fundamental forces, and to any cosmic effects that can change the expansion rate. 
As a result it opens a window to new physics, probing new cosmic components present in the first seconds to minutes at a temperature scale between 10 keV and 10 MeV. As such, it provides one of the deepest probes into the Universe which is based on 1) known physics, and 2) observations. Its success 
using Standard Model inputs (of cosmology and nuclear and particle physics), allows for very little deviation from the Standard Model and as such it is possible to place strong constraints on physics beyond the Standard model which is effective at temperatures around 1 MeV
\cite{subir,cfos,Pospelov:2010hj,Mangano:2011ar,Nollett:2011aa,ysof}. These include dark matter and dark energy, as well as new forms of radiation.
Most notable among these is the limit on the number of relativistic degrees of freedom present during BBN, often parameterized by the number of neutrino flavors. In a recent analysis this was found to be $N_\nu < 3.180$ (95 \% CL) \cite{ysof}. 

While additional relativistic degrees of freedom contribute to the total energy density, and speed-up the expansion of the Universe, they do not change the equation of state. 
In many theories of physics beyond the Standard Model, a species $X$ becomes non-relativistic during BBN, then later (necessarily if it has any effect on BBN) decays or annihilates.
The effect of a non-relativistic particle cannot be described by $N_\nu$, because the cosmic equation of state during BBN is altered \cite{ktw,Kaplinghat:1998wc,Kaplinghat:2000zt,Carroll:2001bv} and thus  requires a dedicated analysis, which is the goal of this paper.
Initial work on this possibility was considered in a pair of papers \cite{st1,st2} which considered the effect of decaying particles on BBN with either entropy producing or inert decays. This question was more recently examined in the context of decaying scalar fields present at the time of BBN \cite{ac} using Alter-BBN \cite{alter}. 
In addition recently the effects of a non-standard expansion rate on BBN was considered in \cite{AristizabalSierra:2023bah}.
The question of equilibration of a new species during or after BBN was considered in \cite{Berlin:2019pbq}.
BBN limits to some specific models were recently applied in \cite{Serpico:2004nm,Depta:2019lbe,Chang:2024mvg} and the effects of entropy injection between BBN and recombination were considered in
 \cite{Sobotka23a} and \cite{Sobotka23b}.

Here we take a fresh look at the question of how large a matter component with equation of state parameter $w = p/\rho = 0$ can be present during BBN. In particular, we make use of recently constructed BBN likelihood functions \cite{foyy,ysof} convolved with Planck likelihood chains \cite{Planck2018}. In standard BBN cosmology, the universe is dominated by radiation, and the time-temperature relation is well known
\beq
\frac{1}{2t} = H = \frac{\rho_{\rm r}^{1/2}}{\sqrt{3}M_P} = \left(\frac{\pi^2 g_{\rm r}}{90} \right)^{1/2} \frac{T^2}{M_P} \, ,
\label{tT}
\eeq
where $H = \dot{a}/a$ is the Hubble parameter, $a$ is the cosmological scale factor, $\rho_{\rm r}$ is the total energy density in radiation, $g_{\rm  r}$ is the number of relativistic degrees of freedom, and $M_P = (8\pi G_{\rm N})^{-1/2}$ is the reduced Planck mass. This leads to the convenient relation $t_{\rm s} T^2_{\rm MeV} = 2.41/\sqrt{g_{\rm r}}$ with $t$ measured in seconds and $T$ in MeV. These relations assume a radiation dominated universe with $w = 1/3$. The presence of a matter component beyond the standard model would alter the equation of state and the Friedmann equation becomes
\begin{equation}
H^2 = \frac{\rho}{3 M_P^2}
    = \frac{\rho_r + \rho_X }{3M_P^2} \, ,
    \label{friedmann}
\end{equation}
where $\rho_X$ is the energy density of the matter component. If $X$, with equation of state parameter, $w$, comes to dominate the energy density, we have 
 $H = \frac{2}{3(1+w)}t$.

Given the change to the cosmological evolution, we recompute the production of the light elements \he4 and D/H \footnote{We also compute the abundances of \he3 though there currently no direct method to connect observations \cite{he3} to primordial abundances \cite{he32}. We also calculate the abundance of \li7, but here too, the observed abundances \cite{sbordone2010,Bonifacio:2012vp,Aguado:2019egq,pinto} may not be representative of primordial abundances \cite{fdo9}.}. Any amount of matter present at the time of BBN which plays a significant role in the expansion history must decay (or annihilate) in order to avoid over-closing the Universe. As such, depending on the final state decay products, either the baryon-to-photon ratio, 
\beq
\eta = \frac{n_{\rm baryon}}{n_\gamma} \  \propto \ (aT)^{-3} \ \ ,
\eeq
and/or the effective number of neutrinos may differ between the time of BBN and 
CMB decoupling. Limits on the change in these quantities was recently considered in \cite{ysof}. It is also possible that 
the decay products directly affect the abundances through post-BBN processing \cite{Kawasaki:1994af,kkm,cefo,jed1,kkm2,kkmy,jp,ceflos,serp,Kawasaki:2017bqm,Hufnagel2018,Kawasaki:2020qxm,Depta2021,Alves:2023jlo}. However, we will assume that the decays of $X$ are not hadronic and we will only consider lifetimes shorter than $10^4$ s when electromagnetic decay products are scattered sufficiently to prevent nonthermal photodissociation of any light nuclides.  For longer lifetimes, these nonthermal effects can lead to stronger constraints \cite{cefo,Hufnagel2018,Kawasaki:2020qxm,Depta2021}.

As noted above, we make use of BBN likelihood functions (the BBN theory likelihood convolved with the observational likelihood) and convolve this with the CMB likelihood functions taken from Planck \cite{Planck2018}. As a result we obtain a global likelihood function $\mathcal{L}(\eta,\tau_X,\xi)$, which is a function of the our three independent input parameters. Here, $\tau_X$ is the lifetime of the matter component,
and $\xi = \rho_X/\rho_{\rm r}$. The total likelihood can then be marginalized to produce
either single or combined likelihood functions on any combination of these parameters. 

In what follows, we develop our formalism in Section \ref{formal}. There we define our parameter space and describe how these are implemented in our BBN calculations. 
In addition, we discuss the effect of decays on changes in $\eta$ and the effective number of neutrino flavors, $N_\nu$. The observations used to construct the likelihood functions are reviewed in Section \ref{sect:inputs}. In Section \ref{results}, we present the results of our BBN calculations and the resulting likelihood function $\mathcal{L}(\eta,\tau_X,\xi)$ for both cases of electromagnetic and invisible decays. 
A summary and comparison with previous results are given in Section \ref{summary}.

\section{Formalism}
\label{formal}

Standard BBN is usually defined as the theory predicting the light element abundances using {\em standard} nuclear physics, {\em standard} particle physics (implying $N_\nu = 3$) and {\em standard} cosmology. In this standard model, 
it is assumed that the Universe is described by an FRW metric and that 
the Universe is radiation dominated with
\begin{equation}
\rho_r = {\pi^2 \over 30} \left( 2 + {7 \over 2} + {7 \over 4}N_\nu \right) T^4 ,
\label{rho}
\end{equation}
at temperatures $T \gtrsim 1$ MeV. The three contributions correspond to photons, $e^\pm$, and neutrinos. The cosmological scale factor scales as $a \propto t^{1/2}$ and $\rho_r$ scales as $a^{-4}$ leading to the time-temperature relation mentioned above. The time-temperature relation $dT/dt$ is found
from combining the Friedmann equation (\ref{friedmann}) with the conservation of the energy momentum tensor, giving
\beq
\dot{\rho} + 3 (1+w) H \rho = 0 \, .
\label{eq:2ndLaw}
\eeq
For a radiation dominated universe, $w=1/3$ and we readily obtain $\rho \propto T^4 \propto a^{-4}$.

While there are many ways to go beyond the standard model (any of them), one way to modify the standard model is to introduce a new component, $X$, with an equation of state parameter, $w_X$. 
As noted above, the presence of a new species $X$ affects BBN  via the cosmic expansion rate as in Eq.~(\ref{friedmann}).
If our species $X$ is decoupled, then we have
separate conservation of the energy-momentum tensor of $X$ so that 
${\dot \rho}_X = 3(1+w_X) H \rho_X$
and this means that terms in $X$ drop out of eq.~(\ref{eq:2ndLaw}). 
This is easily solved and the energy density of $X$, $\rho_X$ scales as
\begin{equation}
\label{eq:rho_X}
    \rho_X(a) = \rho_1 \ \left( \frac{a_1}{a} \right)^{3(1+w_X)} \, ,
\end{equation}
where we set the normalization by specifying $\rho$ to be $\rho_1$ at $a = a_1$. 

  However unless the equation of state parameter, $w_X =1/3$,  the time-temperature relation is altered and two components of the energy density are coupled (through gravity) in the Friedmann equation (\ref{friedmann}). 
When $X$ represents additional light neutrino species (for which $w = 1/3$), then the time-temperature relation is not affected, as also holds for the contributions from Standard Model neutrinos except for small heating effects.
Here however we will consider cases $w_X \ne 1/3$, and so
we can not in general simply recast an arbitrary beyond the standard model component in terms of additional neutrino flavors.

It will be convenient to set the normalization before nucleosynthesis begins, and we choose the time when the (standard model) radiation density is at a temperature of $T_1 = 10$~MeV. 
In other words, we can define a relative 
density parameter by
\beq
\rho_X = \xi \left(\frac{T_1}{T} \right)^{(1-3w_X)} \rho_r \, ,
\eeq
where
$\xi$ is the fraction of the SM radiation density in $X$ at the normalization temperature, $T_1$.
From here onward, we will consider $X$ to be in the form of a pressureless gas, and so we take $w_X=0$.

\begin{figure}[!htb]
 \centering
\includegraphics[width=0.95\textwidth]{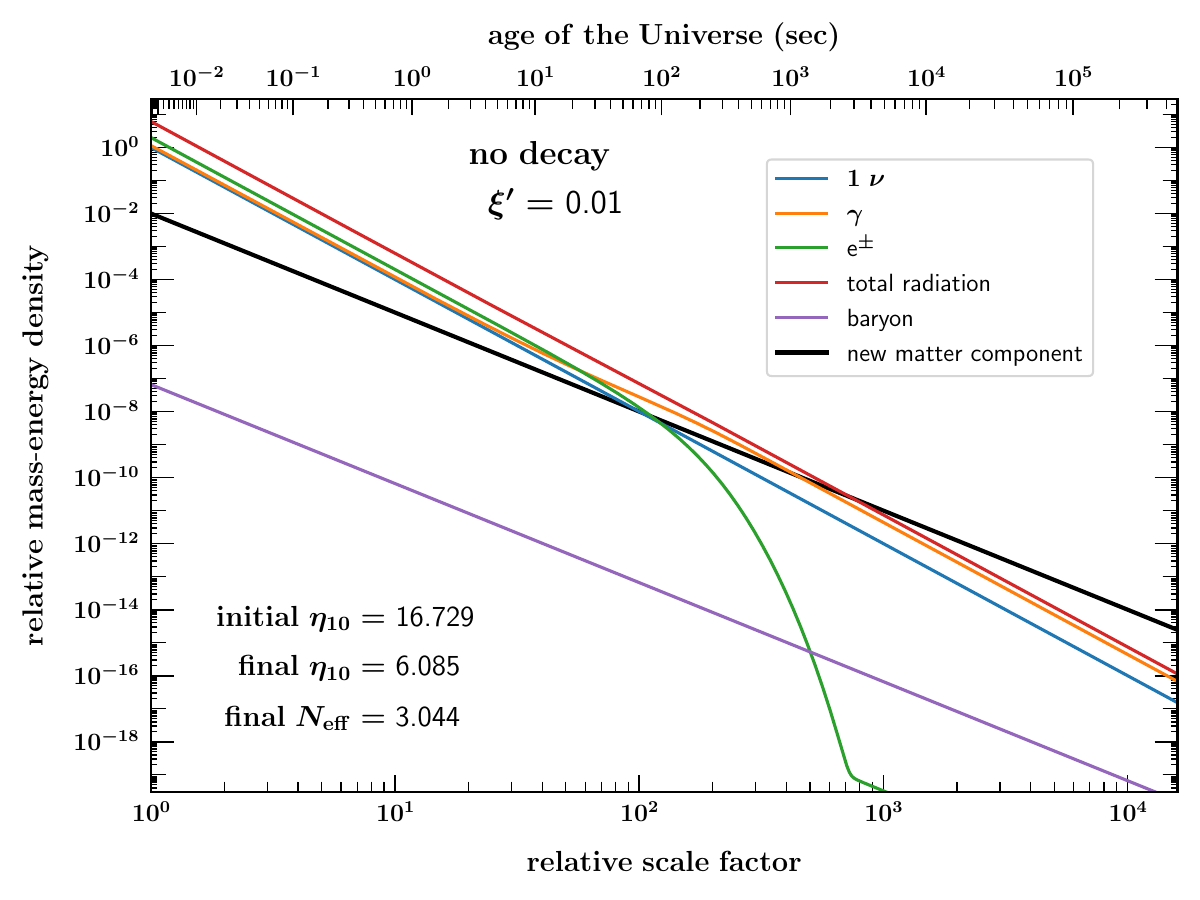}
    \caption{ The evolution of the energy densities of baryons (purple line), $e^\pm$ (green curve), photons (orange line), a single neutrino species (blue line), total radiation (red line), and a new matter component (black line). The $X$-energy density at $T=T_1 = 10$ MeV was set at $\xi' = 0.01$.  }
    \label{fig:evol-enden}
  \end{figure}

In Fig.~\ref{fig:evol-enden}, we show the evolution of several components of the energy density as a function the scale, normalized so that $a  = 1$ at $T = 10$ MeV and the relative energy density is normalized so that the energy of a single neutrino is unity at $a=1$. The baryon density (purple line) which evolves as $\rho_{\rm B} \propto a^{-3}$ is unaffected by the other components. Similarly the new component with $w_x = 0$ and $\rho_X \propto a^{-3}$ (black line) runs parallel to $\rho_{\rm B}$. At early times and on the scale displayed, the energy density in photons, $\rho_\gamma$ (orange line) and the density in a single neutrino flavor (blue line) are very close (they differ by a factor of 7/8) and scale as $a^{-4}$. The $e^\pm$ energy density (green curve) is 7/4 times larger than the photon density at early times. But at a temperature $T \simeq m_e$ corresponding to $t \approx 2.8$~s and $a \approx 20$, $e^\pm$ annihilation is not accompanied by pair production and the $e^\pm$ begins to drop off exponentially, until all of the positrons are annihilated at $t \sim 10^3$ sec.  Thereafter, the density of the remaining electrons (now non-relativistic) is simply parallel to that of the baryons whose charge they balance. As the neutrinos are largely decoupled at this point, 
the energy density released by the annihilations goes into heating up the photons, and one clearly sees that at late times $\rho_\gamma = (8/7)(T_\gamma/T_\nu)^{4/3} \rho_\nu$, where $(T_\nu/T_\gamma)^3 = 11/4$ due to entropy conservation. This same factor is responsible for diluting the baryon-to-photon ratio given in the figure. The new matter component has been normalized to a single neutrino so that $\xi' \equiv \rho_X/\rho_{1\nu} = 0.01$, and exceeds the photon energy density when $T_\gamma \approx (43/22) \xi T_1$.  As one can see, at some point, $X$ must decay or it will continue to dominate the energy and greatly over-close the Universe. As we will see in the next section, this value of $\xi$ was chosen as it is close to the limit imposed by the BBN calculated abundances.

It will be convenient to separate the radiation density $\rho_r = \rho_{e\gamma} + \rho_\nu$.
Here the electromagnetic component of the radiation density, namely photons and $e^\pm$,  is  $\rho_{e\gamma} = \rho_\gamma + \rho_{e^\pm}$,\footnote{In addition, one may include the baryon density, but because $\eta \sim 6 \times 10^{-10}$ it can be safely neglected.} and $\rho_\nu = 3 \rho_{1\nu}$, where the latter is the energy density of a single Standard Model neutrino, $\rho_{1\nu} = (7\pi^2/120) T_\nu^4$ and $T_\nu = (4/11)^{1/3} T$ is the neutrino temperature well after $e^\pm$ annihilation is complete, ie., $T \ll m_e$.
The ratio of the $X$-matter density to a single neutrino, $\xi'$ was defined above, and the ratio to the electromagnetic component
$\rho_{e\gamma}$ is $\xi'' = \rho_X/\rho_{e\gamma}$, both defined at $T = T_1$. 
For convenience we summarize these measures of the $X$ density:
\begin{eqnarray}
    \xi & = & \left. \frac{\rho_X}{\rho_{\rm rad}} \right|_{10 \, \rm MeV} \, , \\
    \xi^\prime & = & \left. \frac{\rho_X}{\rho_{1\nu}} \right|_{10 \,  \rm MeV}  = \frac{43}{7} \xi \, ,\\
    \xi^{\prime \prime} & = & \left. \frac{\rho_X}{\rho_{\gamma}} \right|_{10 \, \rm MeV} = \frac{43}{22} \xi = \frac{7}{22} \xi^\prime\, ,
\end{eqnarray}
all of which are evaluated at $T_1 = 10 \ \rm MeV$, which we choose to ensure that neutrinos are fully coupled to the plasma.
In principle the total density should include dark matter and dark energy, but 
neither should contribute substantially at the time of BBN.

As we have just seen, the new species must decay in order to avoid over-closing the universe today (and starting structure formation too early). 
For example, today, the fraction of the energy density in $X$, relative to the critical density, $\rho_c \equiv 3 H^2 M_P^2$ can be written as
\beq
\Omega_X = \frac{\rho_X}{\rho_c} =  \xi'' \frac{T_1}{T_0} \Omega_\gamma = 2.3 \times 10^6 \ \xi'' \, ,
\eeq
where we have taken the present temperature of the CMB to be $2.3 \times 10^{-10}$ MeV and $\Omega_\gamma = 5.38 \times 10^{-5}$. Thus, unless $\xi''$ is so small that it does not affect BBN, $X$ must decay.
We quantify this via a particle lifetime $\tau_X$, and associated decay rate 
\begin{equation}
    \Gamma_X = \frac{1}{\tau_X} \, .
\end{equation}
Allowing for decays, the equation for energy-momentum conservation is now 
\begin{equation}
\label{eq:drhox/dt}
    \frac{d}{dt} \rho_X + 3 \frac{\dot{a}}{a} \rho_X = -\Gamma_X \rho_X \, ,
\end{equation}
and thus $\rho_X$ evolves as 
\begin{equation}
\label{eq:rhoxvt}
    \rho_X = \rho_1 \pfrac{a_1}{a}^3 e^{-\Gamma_X t}
\end{equation}
showing a factorization of the cosmic volume dilution and the exponential decay.

Decays will occur when the decay rate becomes comparable with the Hubble rate. It is convenient to distinguish between two possibilities: decays occur when $\rho_r > \rho_X$
or when $\rho_r < \rho_X$. In the former case, we can assume that the energy density driving the expansion is dominated by $\rho_r$, and we find the temperature of the radiation when decay occurs from setting $\Gamma_X = 2 H$
leading to 
\begin{eqnarray}
\label{eq:Td}
T_d & = & \left( \frac{90}{4\pi^2 g_d} \right)^{1/4} \left( \Gamma_X M_P \right)^{1/2}
  =  0.86 \ {\rm MeV} \ \pfrac{10.75}{g_d}^{1/4} \ \pfrac{1 \ \rm sec}{\tau_X}^{1/2}
\, ,
\end{eqnarray}
where $g_d = g(T_d)$ is the number of degrees of freedom in the radiation bath at temperature $T_d$. 
While our results will not depend explicitly on the mass of $X$, we will assume that $m_X > T_d$ and all decays -- even those to visible matter -- will occur out of equilibrium.  Clearly to satisfy $\rho_X < \rho_r$ at $T_d$, we must have $T_d > \xi T_1$ and 
\beq
\tau_X < \left(\frac{90}{4\pi^2 g_d} \right)^{1/2} \frac{M_P}{\xi^2 T_1^2} \qquad \rho_X < \rho_r \; @ \; T=T_d \, .
\label{gr}
\eeq
Alternatively, if $X$ decays while it dominates the energy density, decay occurs when $\Gamma_X = \frac32 H$ and 
\beqar
T_d & = & \left( \frac{40}{\pi^2 g_d \xi T_1} \right)^{1/3} \left( \Gamma_X M_P \right)^{2/3}  
 =  0.46 \ \xi^{-1/3} \ {\rm MeV} \ \pfrac{10.75}{g_d}^{1/3} \ \pfrac{1 \ \rm sec}{\tau_X}^{2/3}
\, .
\eeqar
In this case,  to satisfy $\rho_X < \rho_r$ at $T_d$, we must have $T_d < \xi T_1$ and
\beq
\tau_X > \left(\frac{40}{\pi^2 g_d} \right)^{1/2} \frac{M_P}{\xi^2 T_1^2} \qquad \rho_X < \rho_r \; @ \; T=T_d \, .
\label{gm}
\eeq
Note the difference in the right hand side of Eqs.~(\ref{gr}) and (\ref{gm}) is entirely due to 
assuming either radiation dominated in the former or matter dominated (by $X$) in the latter, and as such both are approximations.

\begin{figure}[!ht]
  \centering
\includegraphics[width=0.75\columnwidth]{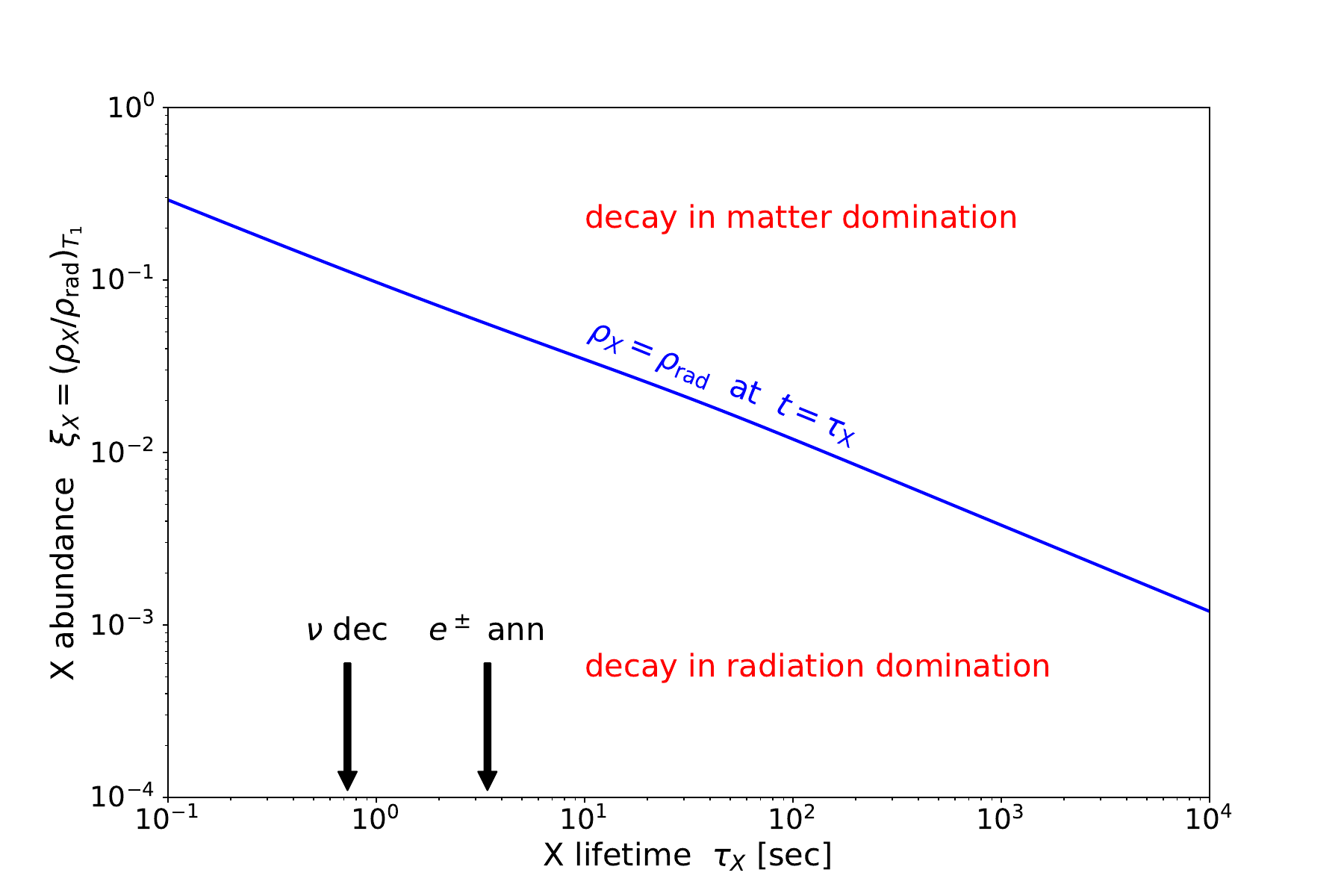}
  \caption{\it 
 The $\tau_X-\xi$ parameter plane. Points above the lines have decays which occur when dominated by $X$, and points below have decays in a radiation dominated era.
  The two arrows are determined by assuming a neutrino decoupling temperature $T_{\nu \rm dec} \simeq 1$ MeV, and a temperature for $e^\pm$ annihilation, $T_e \sim m_e$.  
} 
  \label{fig:tau-xi}
\end{figure}

All of the constraints on the matter domination can be expressed in terms of the two parameters, $\tau_X$ and $\xi$ which characterize the model. The choice of parameters which distinguish whether the decay of $X$ takes place in a matter dominated or radiation dominated is shown in Fig.~\ref{fig:tau-xi}.  At the moment of matter and radiation equality,
 $H = 4(\sqrt{2}-1)/3t$.  In Fig.~\ref{fig:tau-xi}, we show the ($\tau_X, \xi$) parameter plane. 
 The sloped line shows the value of $\xi$ when $\rho_X = \rho_r$ at $T=T_d$ 
 or equivalently $t = \tau_X$. For values of $\xi$ above the line decays occur when the Universe is dominated by the energy density of $X$, and values of $\xi$ below the line have decays in a radiation dominated era. While subtle, the line is not perfectly straight and is bumped up when $e^\pm$ annihilation occurs changing the number of degrees of freedom. The vertical arrows indicate the approximate times corresponding to neutrino decoupling (taken here to be $T = 1$ MeV) and $e^\pm$ annihilation at $T = m_e$.

Whether or not $X$ decays in a matter or radiation dominated period, entropy will be produced and will affect the baryon-to-photon ratio and the effective number of degrees of freedom. Thus either one or both of these quantities may differ at the BBN and CMB decoupling epochs. 
We consider two cases that lead to different results:
\begin{itemize}
\item{\bf Case (a):} $X$ decays to photons or other electromagnetically interacting particles, and

    \item{\bf Case (b):}  $X$ decays to neutrinos or other weakly (or superweakly) interacting particles.
\end{itemize}
For case (a), we can approximate the increase in entropy from electromagnetic decays by considering the resulting reheat temperature defined by
\beq
\frac{g_d \pi^2}{30} T_R^4 = \frac{g_d \pi^2}{30}T_d^4 + \rho_X(T_d) \, ,
\eeq
and the entropy increase is therefore
\beq
\left(\frac{T_R}{T_d}\right)^3 = \left(1+\frac{\xi'' T_1}{T_d} \right)^{3/4} \, .
\eeq
In this case, the value of $\eta$ is reduced by a factor $(T_d/T_R)^3$. 
\begin{eqnarray}
\label{eq:etachange}
    \eta  & = & \eta_1 \left(1+\frac{\xi'' T_1}{T_d} \right)^{-3/4} \, ,
\end{eqnarray}
where $\eta_1 = \eta_{10}(T_1)$ is the initial baryon-to-photon value.\footnote{For convenience, $\eta_{10} \equiv \eta \times 10^{10}$.}
We see that the decays act to {\em decrease} $\eta$.  We will use the CMB likelihood to constrain the {\em final} baryon-to-photon value.  Thus as  $\xi^{\prime \prime}$ increases, this requires a higher {\em initial} $\eta_1$.  This will have important consequences for the light-element abundances.

For the electromagnetic decays of case (a), the number of effective neutrinos may also be affected, depending on the temperature at which the decay occurs.
For decays which occur at a temperature $T> T_{\nu \rm dec}$, $N_\nu$ is unaffected because
photons share their the decay energy and entropy with neutrinos. 
But for $T< T_{\nu \rm dec}$, decays heat the photons but not the neutrinos, so that $T_\nu/T_\gamma$ is lower than in the standard case.  The result is that $N_\nu$ is {\em reduced} by a factor of $(T_d/T_R)^4$.
Thus
\begin{align}
    N_\nu  & =  3 & &  N_{\rm eff} = 3.044 & {\rm for}~~T_d > T_{\nu \rm dec} \, , \label{a1}\\
    N_\nu & =  3  \left(1+\frac{\xi'' T_1}{T_d} \right)^{-1} && N_{\rm eff} = 3\left(1+\frac{\xi'' T_1}{T_d} \right)^{-1} + 0.044 & {\rm for}~~T_e < T_d < T_{\nu \rm dec} \, ,\label{a2} \\
     N_\nu & =  3  \left(1+\frac{\xi'' T_1}{T_d} \right)^{-1} && N_{\rm eff} = 3.044\left(1+\frac{\xi'' T_1}{T_d} \right)^{-1} & {\rm for}~~T_d < T_{e} \, ,
     \label{a3}
\end{align}
 and we assume $N_\nu = 3$ at $T=T_1$.
It is important to note that we distinguish between $N_\nu$ and the effective number neutrinos $N_{\rm eff}$. Recall that $N_{\rm eff} > 3$ in the standard model as $e^\pm$ annihilation occurs before neutrinos are completely decoupled. Thus in the standard model, $N_{\rm eff} = 3.044$ \cite{Akita:2020szl,Bennett:2020zkv,EscuderoAbenza:2020cmq,Froustey:2020mcq,Cielo:2023bqp}. This is unchanged if 
$T_d > T_{\nu \rm dec}$. If $T_e < T_d < T_{\nu \rm dec}$, then the number of neutrinos is diluted but $e^\pm$ annihilations still provide 4.4\% of a neutrino. Finally when 
$T_d < T_{e}$, it is $N_{\rm eff}$ that is diluted.

For case b) of invisible decays, $\eta$ is unaffected, and
the decays into invisible particles effectively lead to an {\em increase} in $N_{\nu}$ by a factor 
\beq
\frac{\rho_\nu+\rho_X(T_d)}{\rho_\nu} = 1 + \frac{\xi' T_1}{3T_d} \, ,
\label{neudil}
\eeq
if the decay occurs after neutrino decoupling.
If the decay occurs before decoupling, then it matters if decays are to neutrinos, which are not invisible at this point.  For decays into neutrinos no change in  $N_\nu$ occurs, but $\eta$ is reduced as in case (a).  For decays into other invisible species (which act as radiation), $N_{\rm eff}$ is increased by the same factor as in Eq.~(\ref{neudil}). 
Thus for dark (non-neutrino) decays, we have
\begin{align}
     N_\nu & =  3  \left(1+\frac{\xi' T_1}{3T_d} \right) && N_{\rm eff} = 3.044 + \frac{\xi' T_1}{T_d}  \, .
     \label{b1}
\end{align}
and we see that $N_\nu$ {\em increases} due to decays, in contrast to the electromagetic case.

\begin{figure}[!htb]
  \centering
     \includegraphics[width=0.475\textwidth]{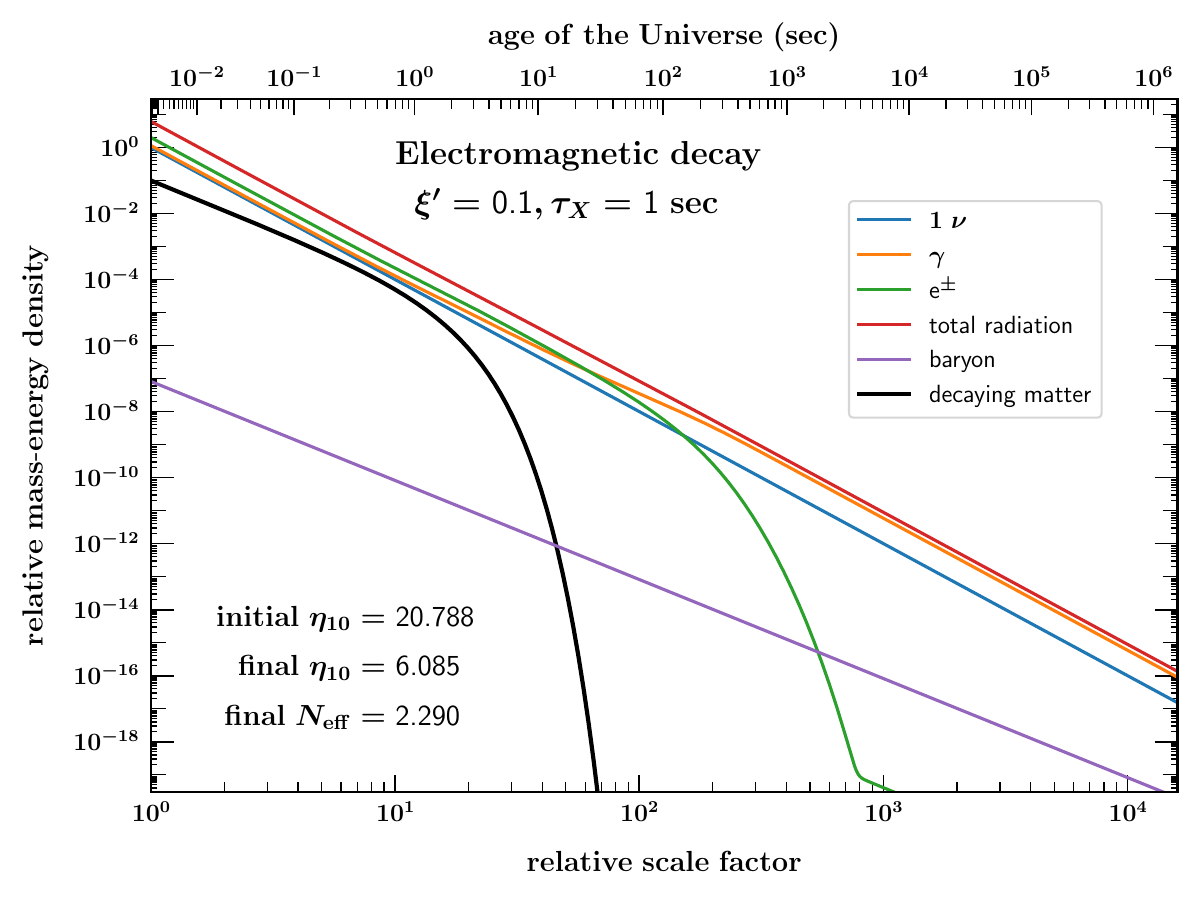}
\includegraphics[width=0.475\textwidth]{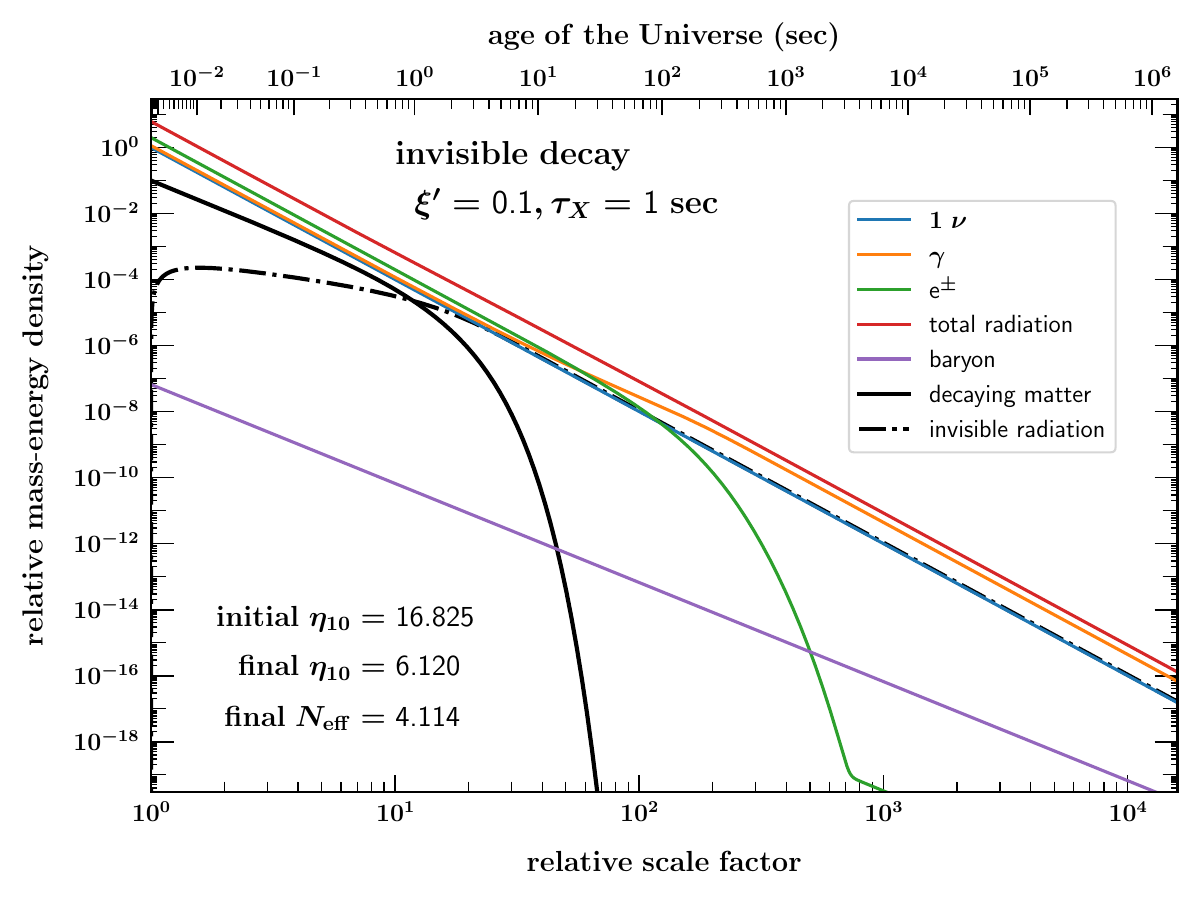} \\
   \includegraphics[width=0.475\textwidth]{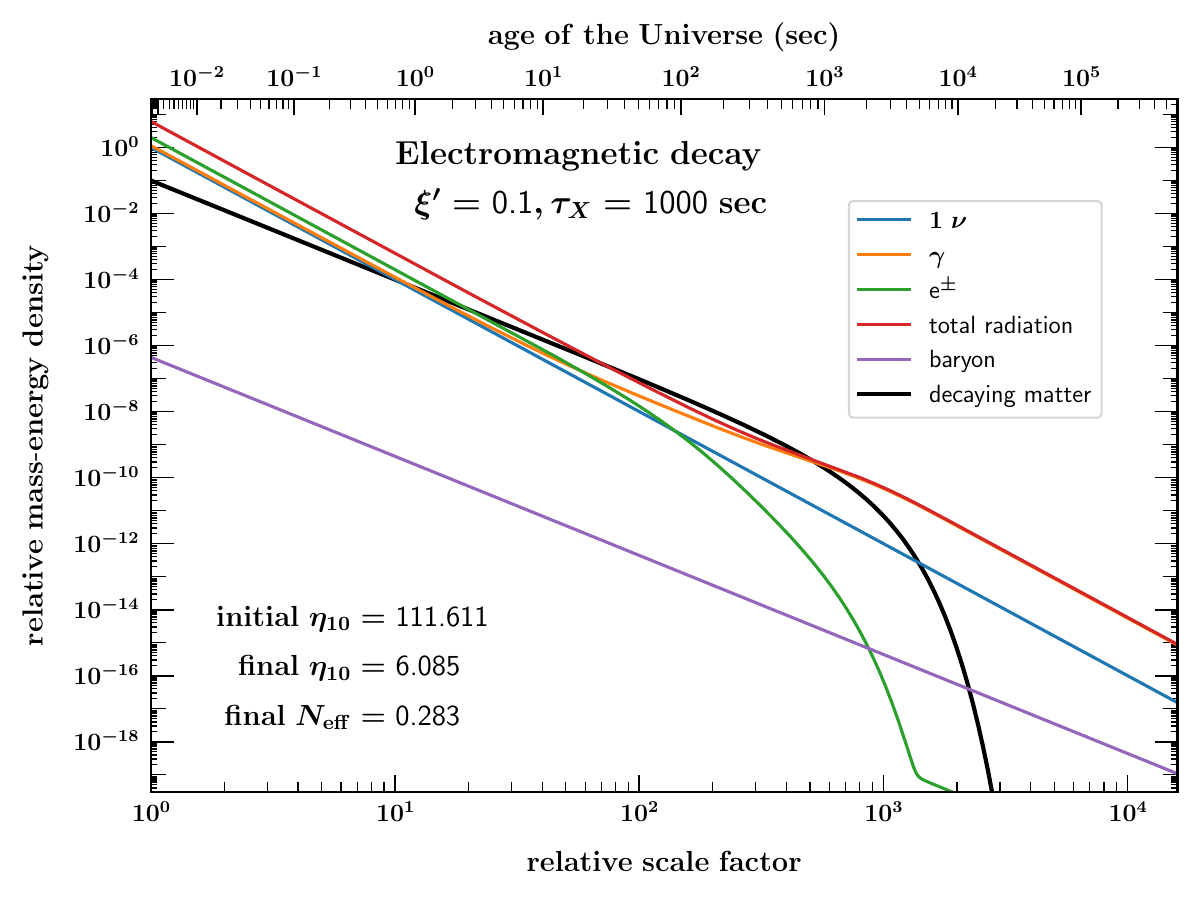}
\includegraphics[width=0.475\textwidth]{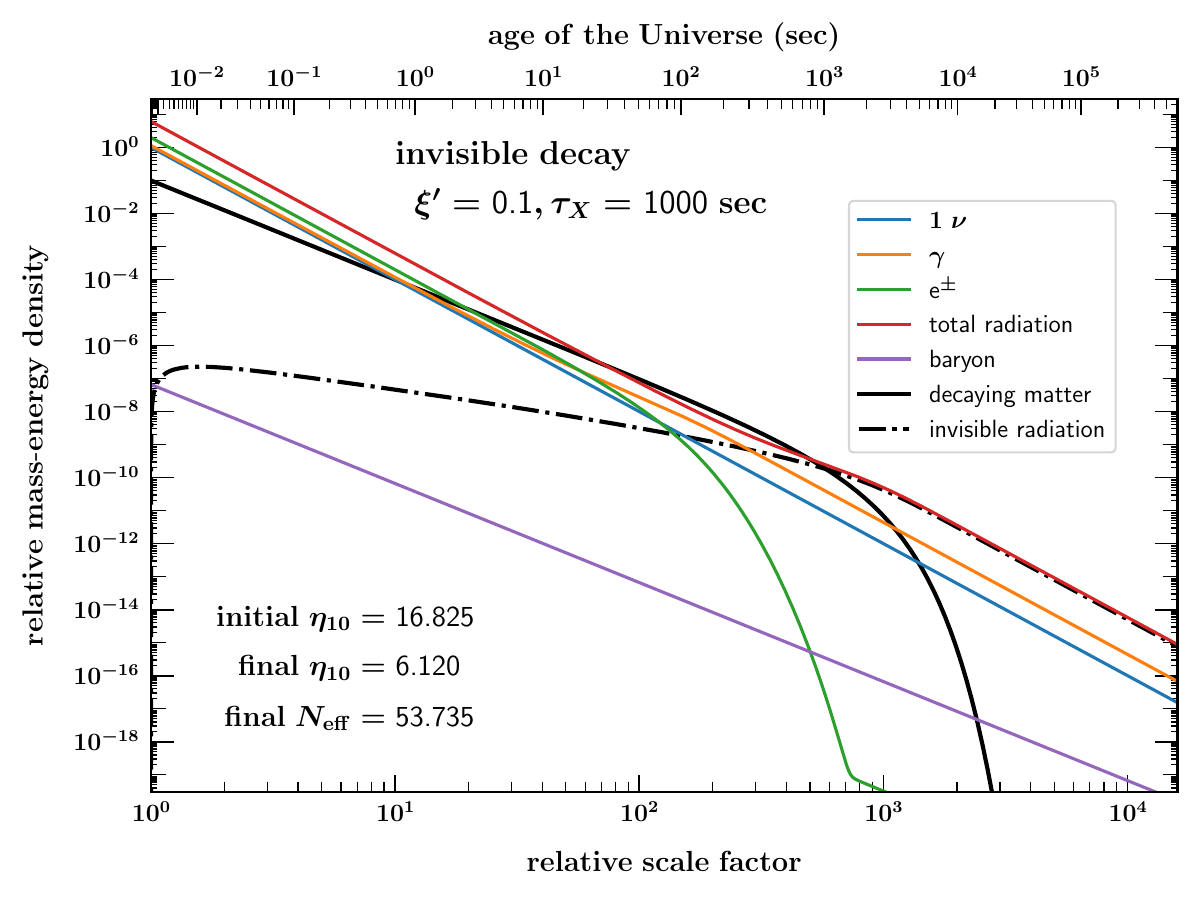}
    \caption{The evolution of the energy densities of baryons (purple line), $e^\pm$ (green curve), photons (orange line), a single neutrino species (blue line), and a new matter component (black line). Here, $\xi' = 0.1$ with $\tau_X = 1$~s (upper) and $\tau_X = 1000$~s (lower). $X$ decays to electromagnetically interacting particle in the left panels and to a dark component in the right panels. The dashed-dotted curve in the right panels shows the evolution of this dark component. Note that the adopted value for the final value of $\eta_{10}$ is different for EM and dark decays as explained in the text. The initial value of $N_\nu = 3$ is fixed in all panels. }
    \label{fig:evol-endendec}
  \end{figure}

The evolution of the energy densities for the two cases is shown in Fig.~\ref{fig:evol-endendec} with case (a) on the left and case b on the right. 
Here, we have taken $\xi' = 0.1$
and $\tau_X = 1$~s (upper), $\tau_X = 1000$~s (lower). For this value of $\xi'$, and we see from Fig.~\ref{fig:tau-xi}, that for the two choices of $\tau_X$,  $X$ decays before it comes to dominate the energy density, i.e., in a radiation dominated universe for $\tau_X = 1$~s and after it starts to dominate in matter dominated universe for $\tau_X = 1000$~s. For case (a), there is a small (though non-zero) effect on the energy density of the electromagnetically interacting particles. For case b, the photon energy density is unaffected, but we see the appearance of a component of dark radiation (dashed-dotted curve). 
At early times, the energy density of the radiation scales as $\rho_{{\rm dark}r} \propto a^{-3/2}$ (rather than $a^{-4}$)\cite{Scherrer:1984fd}
as early decays continuously add to the energy density of the dark radiation. Once the once the lifetime becomes shorter than the age of the Universe, the density $X$ drops exponentially and the density of the dark radiation redshifts as $a^{-4}$ as might be expected.


Note that in each case considered, we adopt an initial value for $N_\nu = 3$. As discussed above, in the absence of a matter component, this leads to a final $N_{\rm eff} = 3.044$. The effects of the matter component may decrease $N_{\rm eff}$ (in case a) - as seen in Eqs.~(\ref{a2}) and (\ref{a3}) or increase $N_{\rm eff}$ (in case b) - as seen in Eq.~(\ref{b1}). The final value of $N_{\rm eff}$ is given in each of the panels in Fig.~\ref{fig:evol-endendec}. Also shown is the final value of $\eta_{10}$. These values are chosen from the likelihood analysis in the next section and the values differ for cases (a) and (b) and will be discussed below. The initial value of $\eta$ ($\eta_1$), also reported in the figure, is related to this final value as approximated by Eq.~(\ref{eq:etachange}) for case (a), and is unchanged in case (b), modulo the standard dilution of $\eta$ due to $e^\pm$ annihilation by the factor of (4/11).

\section{Inputs:  Light Element Abundances and the CMB}
\label{sect:inputs}

The constraints from BBN rely on accurate observational  determinations of the light element abundances, and on cosmological parameters derived from the CMB. These are discussed in detail elsewhere, e.g., \cite{foyy}; we summarize the results here.

Deuterium is observed in high redshift quasar absorption systems \cite{pc,cooke,riemer,bala,cookeN,riemer17,zava,CPS}, where the isotopic abundance is now determined with accuracy of approximately 1\%, giving 
\beq \label{dhobs}
\left(\frac{\rm D}{\rm H}\right)_{\rm obs} = (2.55 \pm 0.03) \times 10^{-5} \, .
\eeq
\he4 is observed in extragalactic HII regions
using a series of \he4 and H emission lines.
The observational determination of \he4
has improved recently \cite{aos4,abopss,Aver:2021rwi,Hsyu:2020uqb,Kurichin:2021ppm,vpp}.
A recent analysis including high quality observations of the Leoncino dwarf galaxy leads to  an inferred primordial abundance of \citep{Aver:2021rwi}
\beq
\label{eq:Ypobs}
Y_{p,\rm obs} = 0.2448 \pm 0.0033 \, .
\eeq
These mean abundances with their 1$\sigma$ uncertainties (assumed Gaussian) allow us to define an observational likelihood function $\mathcal{L}_{\rm obs}$ for \he4 and D/H. As noted earlier, though we compute the abundances of \he3 and \li7, we do not construct observational likelihood functions for these isotopes. 

Turning to the CMB, temperature and polarization anisotropies famously encode a wealth of cosmological parameters,
which depend on the cosmological model assumed.  Since we consider cases where decays can change the cosmic radiation content, we are interested in models in which $N_{\rm eff}$ can vary.
As we showed recently, these models
give a baryon density or baryon-to-photon ratio, $\eta_{10} = 6.090 \pm 0.061$ based on {\em Planck} data alone. 
Furthermore, the value of $N_\nu$
is found to be $N_\nu = 2.800 \pm 0.294$.\footnote{These values differ slightly from the those published in \cite{Planck2018}, which are derived from likelihood chains which assume an a priori relation between the helium abundance and the baryon density. 
} 
The results which include BBN will be discussed below.

\section{Results}
\label{results}

As discussed above, we have modified our BBN code to not only account for an additional component to the energy density, but also to account for the possibility that the new component has an equation of state which differs from $w=1/3$.
This results in a change in the time-temperature relation which is so crucial in the competition between the expansion time-scale and the rate for nuclear reactions. However a few comments are in order before we begin to present our constraints. First, we have noted that we consider the possibility that $X$ decays either into products with electromagnetic interactions (case a), or into dark radiation (case b). 
If $X$ decays into hadronic products which can affect directly the abundances of the light elements during or immediately after BBN, the constraints on $\xi$ are significantly stronger and the question of matter domination during BBN becomes moot \cite{Kawasaki:1994af,kkm,cefo,jed1,kkm2,kkmy,jp,ceflos,serp,Kawasaki:2017bqm}. The change in the energy density and the time-temperature relation become irrelevant. Therefore we do not consider this possibility here. Second, so long as the lifetime of $X$ is sufficiently low ($\tau_X \lesssim 10^4$~s),
the electromagnetic decay products thermalize rapidly due to large radiation density and have a negligible effect on the post-processing of the BBN nuclei.  However, for longer lifetimes, the energies of the decay products is not downgraded and post-process is again significant and very strong constraints on $\xi$ can be derived
\cite{cefo,Hufnagel2018,Kawasaki:2020qxm,Depta2021}. Therefore we do not consider lifetimes $\tau_X \gtrsim 10^4$~s for case (a). 
See for example \cite{kkm2,ceflos} for constraints on decaying particles with both hadronic and electromagnetic decay products.

\begin{figure}[!htb] 
  \centering
  \includegraphics[width=0.475\textwidth]{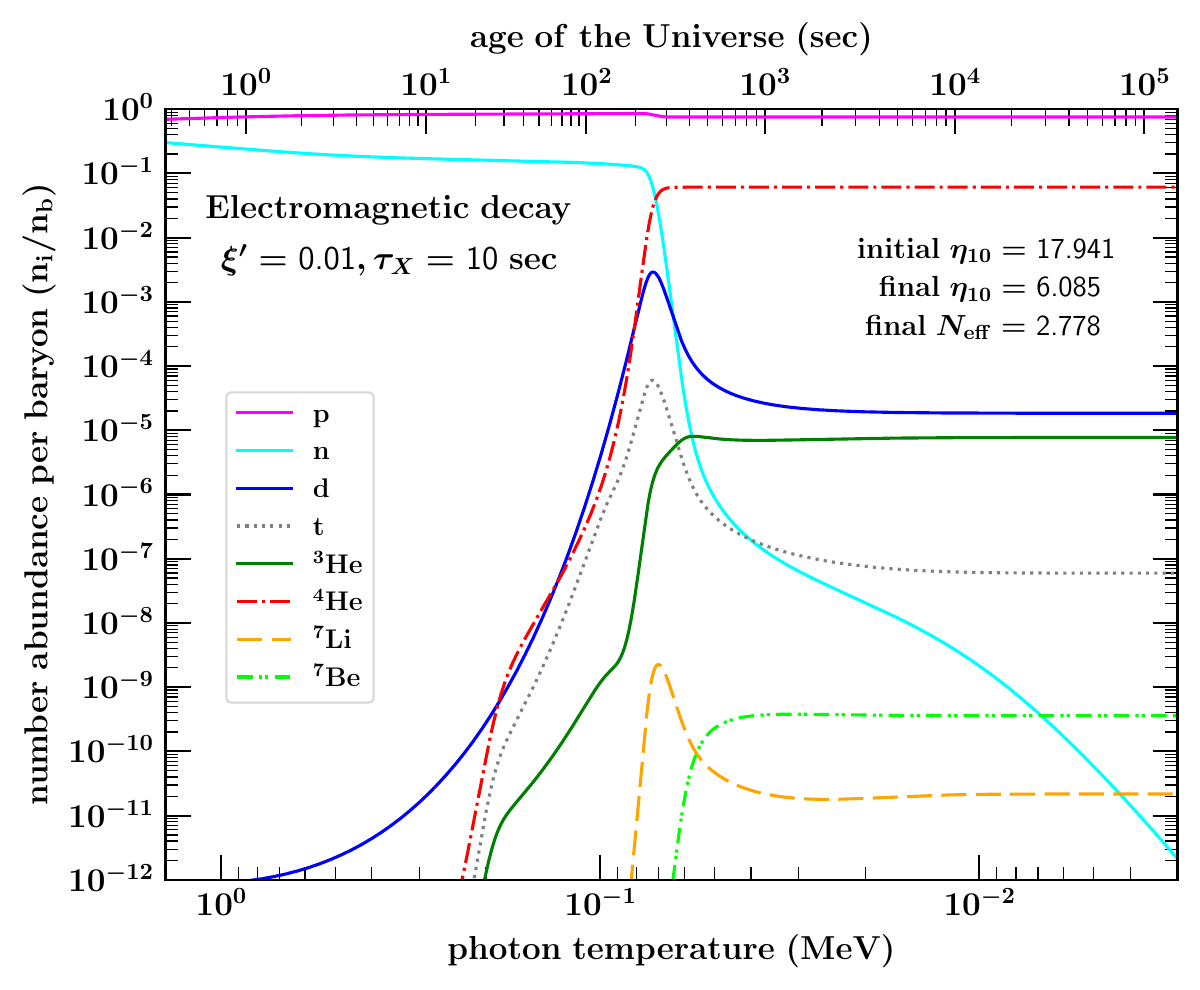}
\includegraphics[width=0.475\textwidth]{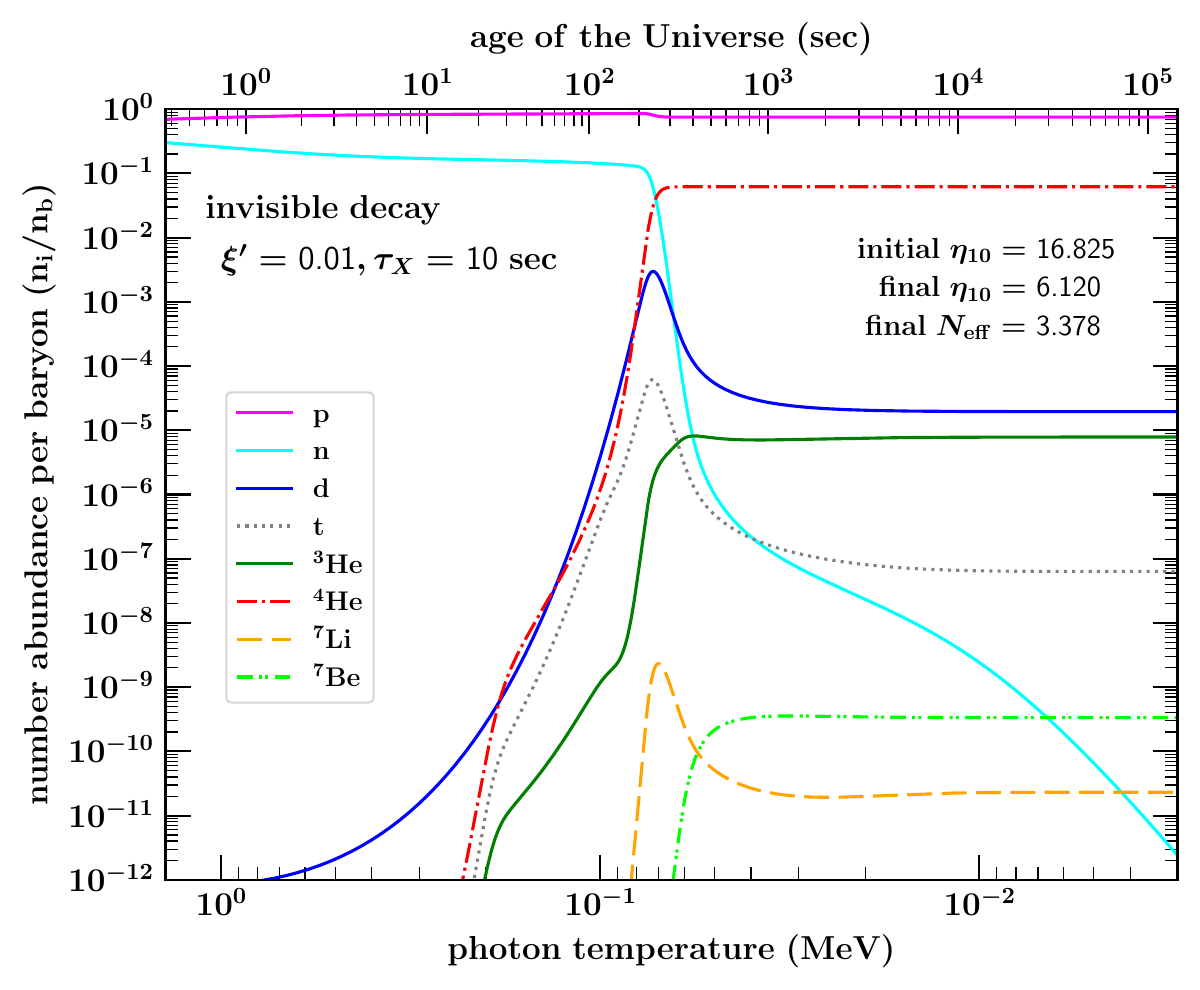}\\
\includegraphics[width=0.475\textwidth]{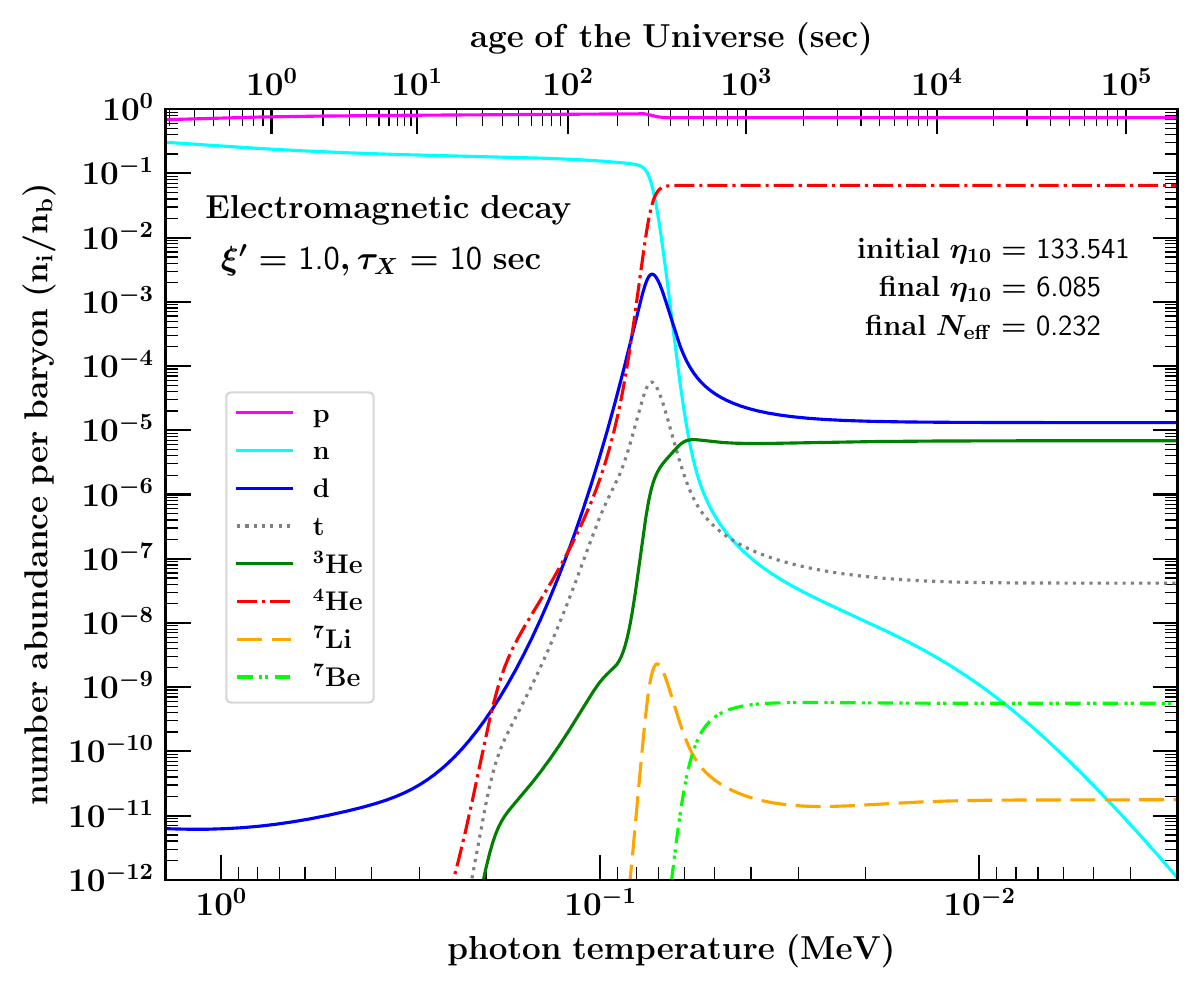}
\includegraphics[width=0.475\textwidth]{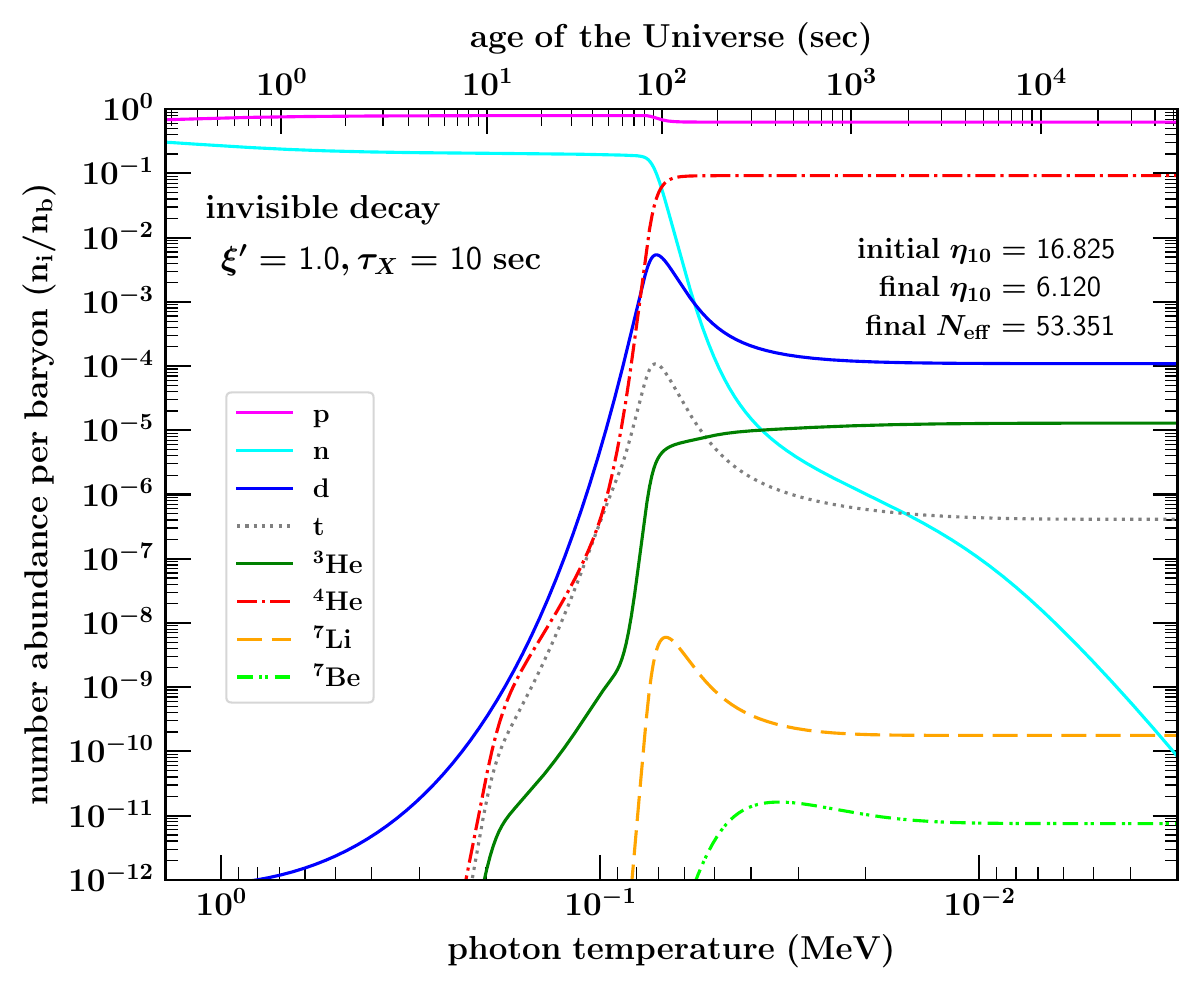}
    \caption{The evolution of the abundances of the light elements produced during BBN with $\tau_X=10$~s and $\xi' = 0.01$ (upper), $\xi' = 1 $ (lower) for EM decays (left) and dark decays (right).  As in Fig.~\ref{fig:evol-endendec}, the adopted final $\eta_{10}$ are different for EM and dark decays. }
    \label{fig:evol-abund}
  \end{figure}

To get an idea of the effect on the light element abundances, we show in Fig.~\ref{fig:evol-abund}
the evolution as a function of time (and temperature) of the baryonic components involved in BBN for $\xi'= 0.01$ (upper panels) and $\xi' = 1$ (lower panels) for EM decays (left panels) and dark decays (right panels) all with $\tau_X = 10$~s (though the effect of $\tau_X$ can not be seen on the scale of these plots).  We see that the rapid decline in the neutron density occurs as the light elements abundances grow. For EM decays, this is delayed when the matter density is high (as in the lower left panel) due to the increased expansion rate. As a consequence, though one can not see it easily on the scale of the plot, the \he4 abundance increases slightly when the $X$ abundance is initially large. This can be traced to the the fact at higher $\xi'$, the initial value of $\eta$ ($\eta_1$) is higher as can be seen from Eq.~(\ref{eq:etachange}). The values of $\eta_1$ are given in the figure. Recall that we have fixed the final value of $\eta_{10} = 6.085$ from the likelihood analysis below. For small $\xi'$ as in the upper left panel, 
the initial value of $\eta_{1} = 17.941$ is similar to the standard initial value
$(11/4)\times 6.085 = 16.733$.
For large $\xi'$ as in the lower left panel, the initial value of $\eta_{10} = 48.5$ and differs significantly leading to the changes in the element abundances. Since the \he4 abundance increases (logarithmically) with $\eta$, this increase in $\eta$ leads to a higher helium abundance.  What is more easily visible is the decrease in the D/H abundance by a factor of about 0.7 as $\xi'$ is increased from 0.01 to 1. This is significant as the observational uncertainty in D/H is about 1\% (see Eq.~(\ref{dhobs})).  This drop can also  be traced to the increased initial value of $\eta$ as D/H decreases with increased $\eta$. The same is true for the \he3 abundance.  
The \li7 abundance increases slightly, as its origin is \be7 whose abundance increases with $\eta$. The final abundance of each of the element isotopes is collected in Table~\ref{tab:abund}. As in Fig.~\ref{fig:evol-endendec},
in case (a) the drop in $N_{\rm eff}$.
This drop is significant when $\xi'$ is large.

\begin{table}[!htb]
\caption{Final abundances of the light element isotopes in each of the cases considered in Fig.~\ref{fig:evol-abund}.
\label{tab:abund}
}
\vskip.1in
\begin{center}
\begin{tabular}{|l||c|c|c|c||}
\hline
 Case & EM:\ $\xi'=0.01$ &  EM:\ $\xi'=1$  & Dark:\ $\xi'=0.01$  &  Dark:\ $\xi'=1$  \\
\hline\hline
\he4 & 0.2461 & 0.2621 & 0.2497 &  0.3713 \\
\hline
(D/H)$\times 10^5$ & 2.44 &  1.78 & 2.60 &  17.3 \\
\hline
(\he3/H)$\times 10^5$ & 1.03 & 0.94 & 1.06 &  2.12 \\
\hline
(\li7/H)$\times 10^{10}$ & 5.05 &  7.83 & 4.77 &  2.94 \\
\hline
\end{tabular}
\end{center}
\end{table}

For completeness, we also show in Fig.~\ref{fig:evol-abund} the abundance evolution when the $X$-decay products are dark. Recall that $\eta_{10}$ is unchanged in this case up to its standard model dilution, however $N_{\rm eff}$ increases and is significantly altered when $\xi'$ is large as in the lower right panel of Fig.~\ref{fig:evol-abund}. This leads to the sizeable changes in the element abundances displayed in the figure. 

To obtain constraints in the ($\eta, \tau_X, \xi $) parameter space we construct likelihood functions for the CMB, the BBN abundances, and as noted above the observational abundances. 
The CMB likelihood, $\mathcal{L}_{\rm NCMB}(\eta,N_\nu,Y_P)$ is taken from  {\em Planck 2018} data. We use the \verb+base_nnu_yhe_plikHM_TTTEEE_lowl_lowE_post_lensing+ chains\footnote{\href{https://wiki.cosmos.esa.int/planck-legacy-archive/index.php/Cosmological_Parameters}{https://wiki.cosmos.esa.int/planck-legacy-archive/index.php/Cosmological\_Parameters}. }
which provide the likelihoods when the number of neutrinos, $N_\nu$, is  allowed to vary. While the BBN abundances are computed here for each value of ($\eta, \tau_X, \xi $), the uncertainties stemming from the nuclear rates are taken from the Monte-Carlo analysis in \cite{foyy,ysof}, and allows us to construct the BBN likelihood $
\mathcal{L}_{\rm mBBN}(\eta,\xi,\tau_X)$. This differs from the standard BBN likelihood as it allows for an extra matter component which affects the expansion rate and the time-temperature relation and thus evolution of the nuclear rates.  As the uncertainty in the abundances are dominated by the uncertainties in the nuclear rates, they are relatively insensitive to the choice of $\xi$, and $\tau_X$.  Our total likelihood function comes from the convolution of the CMB, BBN and observational likelihood functions, 
\begin{equation}
    {\cal L}_{\rm mBBN+NCMB+obs}(\eta, \tau_X, \xi )
    \ \propto \ 
    \int {\cal L}_{\rm NCMB}(\eta,N_\nu,Y_p) \ 
    {\cal L}_{\rm mBBN}(\vec{X};\eta,  \tau_X,\xi) \ 
    \prod  {\cal L}_{\rm obs}(X_i) \ dX_i \ ,
    \label{BBNCMBobs}
\end{equation}
by integrating over the abundances $X_i = Y_P$ and D/H. 
Note that although $N_\nu$ is an argument of ${\cal L}_{\rm NCMB}$, 
it is not an argument of the convolved likelihood function. 
We are assuming $N_\nu = 3$ at the onset of BBN. However, the decays of $X$, may affect the number of relativistic degrees of freedom parameterized as $N_\nu$ as given in Eqs.~(\ref{a1},\ref{a2},\ref{a3}) and (\ref{b1}). This is used as an input to ${\cal L}_{\rm NCMB}$, but it completely determined by $(\tau_X, \xi)$.

Before we present the results of the current analysis, we recall the {\em standard} BBN likelihood results when $N_\nu$ is {\it not} fixed to be 3. When ${\cal L}_{\rm mBBN}(\vec{X};\eta,  \tau_X,\xi)$ is replaced with ${\cal L}_{\rm NBBN}(\vec{X};\eta,N_\nu)$ and marginalized over the abundances of D and He, we obtain the BBN likelihood 
(with arguments $(\eta, N_\nu)$).
This results in a mean (and peak) 
value for $\eta_{10} = 6.088 \pm 0.054$ and a mean value for $N_\nu = 2.898 \pm 0.141$ with a peak value of $2.895^{+0.142}_{-0.141}$ \cite{ysof}.
While this is perfectly consistent with the Standard Model value of 3 (the 95\% CL upper limit is $N_\nu < 3.18)$, it does provide a  slight preference for a {\em decrease} in $N_\nu$ as might be obtained by 
a component of non-relativistic matter as achieved in Eqs.~(\ref{a2}) and (\ref{a3}), when
$\xi \ne 0$ and $T_d < T_{\nu \rm dec}$.

Indeed for case (a), Eqs.~(\ref{a2}) and (\ref{a3}) allow us to estimate a preferred value of
$\xi'' T_1/T_d$. For a given preferred CMB value of $N_\nu$, we can determine
\beq
{\xi_{\rm pref}''} = \frac{3-N_\nu}{N_\nu} \frac{T_d}{T_1} \, .
\eeq
Ignoring the weak dependence on $g_d$, and writing $T_d/1~{\rm MeV} \simeq (\tau_x/1~{\rm s})^{-1/2}$ as in eq.~(\ref{eq:Td}), we have roughly
\beq
\xi_{\rm pref}' \sim 0.01 \pfrac{1~{\rm s}}{\tau_X}^{1/2} \ ,
\label{rest}
\eeq
where we have included the factor of 22/7 to write $\xi'$.
For case b), since we expect an increase in $N_\nu$, the peak of the likelihood function should correspond to the Standard Model value with $\xi' = 0$ and does as we shall see below.

\subsection{Case (a):  Electromagnetic Decays}

\begin{figure}[!ht]
        \centering
        \includegraphics[width=0.475\textwidth]{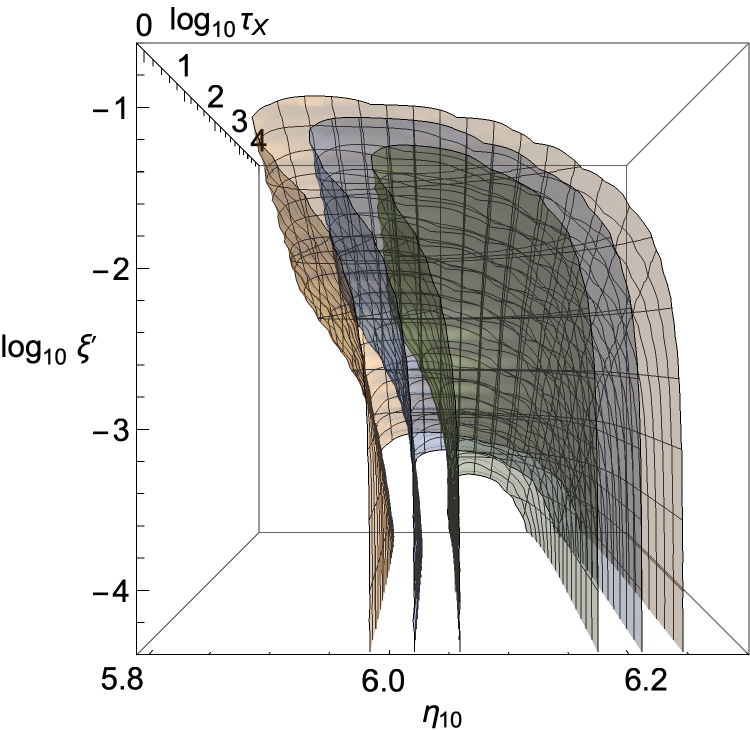}
        \includegraphics[width=0.475\textwidth]{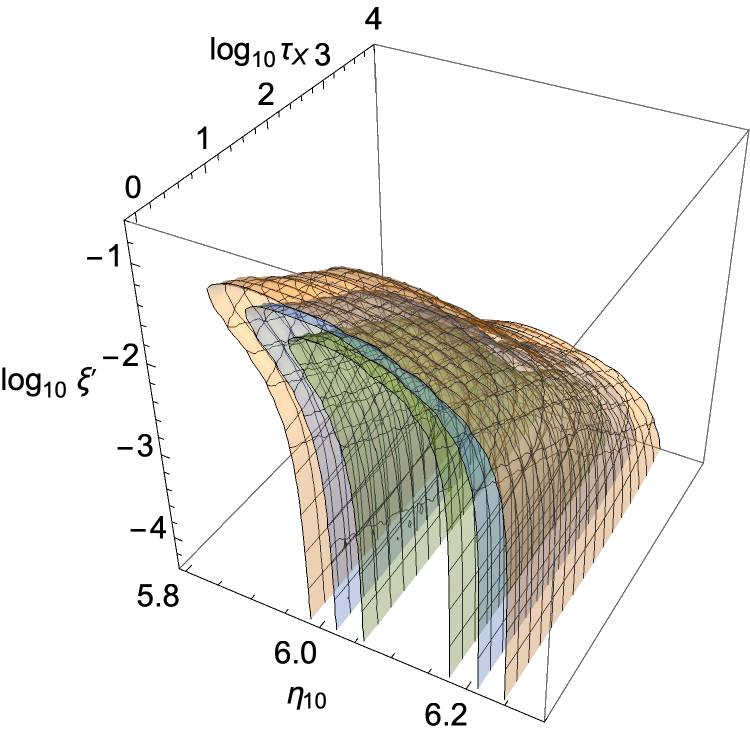}
        \caption{Likelihood contours rendered in the 3-D $(\eta,\tau_X,\xi^\prime)$ space.  We see that the CMB-preferred $\eta$ value leads to a nearly planar region of high likelihood in $(\tau_X,\xi^\prime)$; the walls go straight down for even lower $\xi^\prime$.  There is a slight shift to lower $\eta$ at high $\xi^\prime$, where the likelihood is maximum.}
        \label{fig:EM3D}
    \end{figure}

\begin{figure}[!ht]
        \centering
        \includegraphics[width=0.95\textwidth]{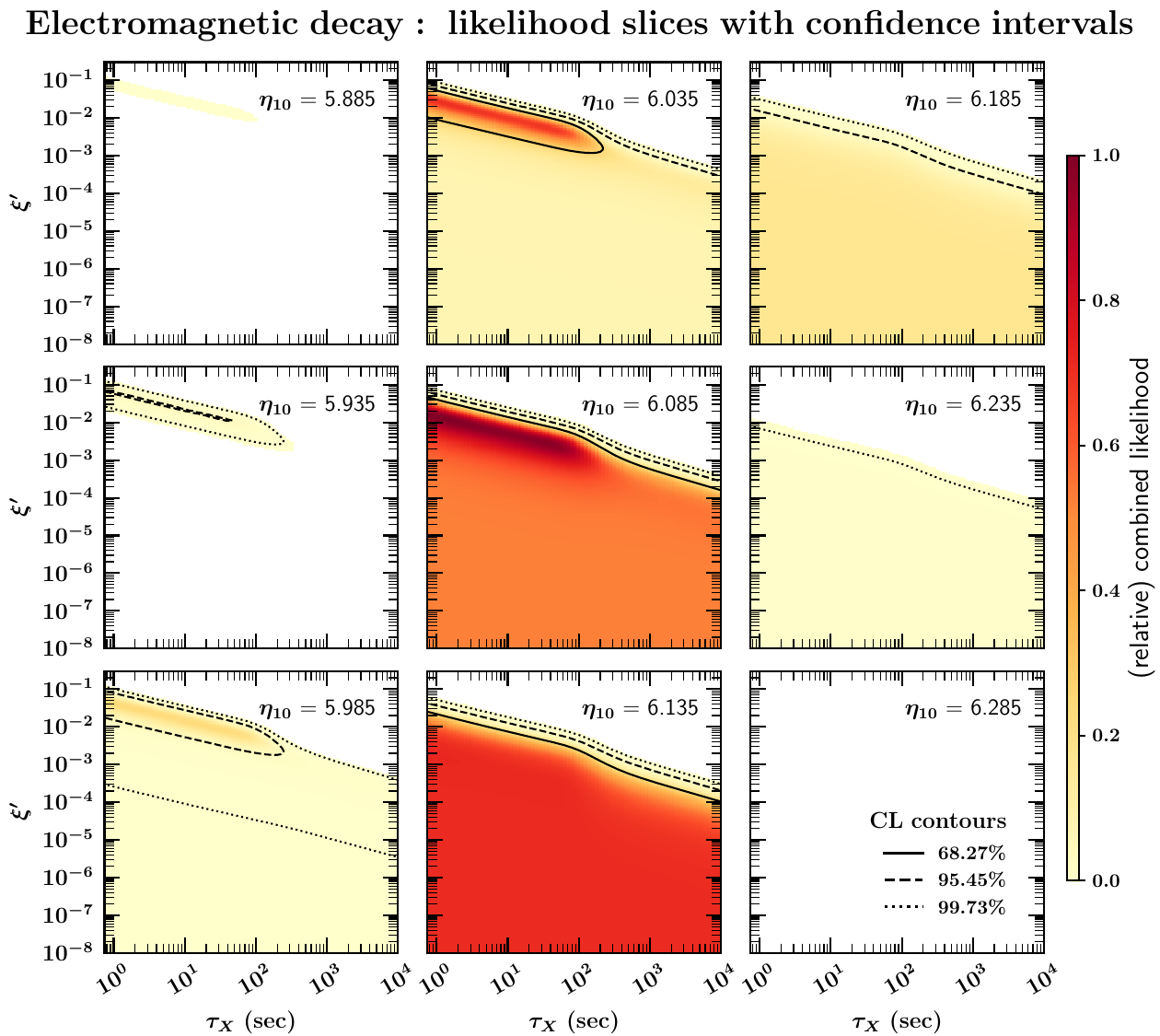}
        \caption{Likelihood contours in the $(\eta, \xi^\prime,\tau_X)$ space, shown as slices at fixed $\eta$. Colors denote the relative value of the likelihood function indicated by the scale at the right. The 68\%, 95\% and 99\% CL contours are shown by solid, dashed, and dotted curves respectively.   }
        \label{fig:EMsliced}
    \end{figure}

We begin by discussing our results for case (a) with visible (electromagnetic) decays. Figure \ref{fig:EM3D} shows likelihood contours rendered in the 3-D
$(\eta,\tau_X,\xi^\prime)$ space.  We see that the contours prefer $\eta$ to be around the CMB {\em Planck} value. This leads to nearly planar features in $(\tau_X,\xi^\prime)$ around a relatively thin range in $\eta$.  The exception is at the largest $\xi^\prime$ values, where the contours curve towards slightly lower values of $\eta$. This can be understood when considering the positive correlation between $\eta$ and $N_\nu$ in the CMB data \cite{Planck2018,ysof}. Since large $\xi'$ reduces $N_\nu$ it must be compensated for by lowering $\eta$.   As we will see, the global maximum likelihood is in this region of somewhat reduced $\eta$. We also see that within this preferred range for $\eta$, the matter abundance is constrained to lower values as its lifetime is increased, as predicted by Eq.~(\ref{rest}). This effect is clearly seen in the likelihood slices shown in Fig.~\ref{fig:EMsliced}. 

\begin{figure}[!ht]
        \centering
        \includegraphics[width=0.8\textwidth]{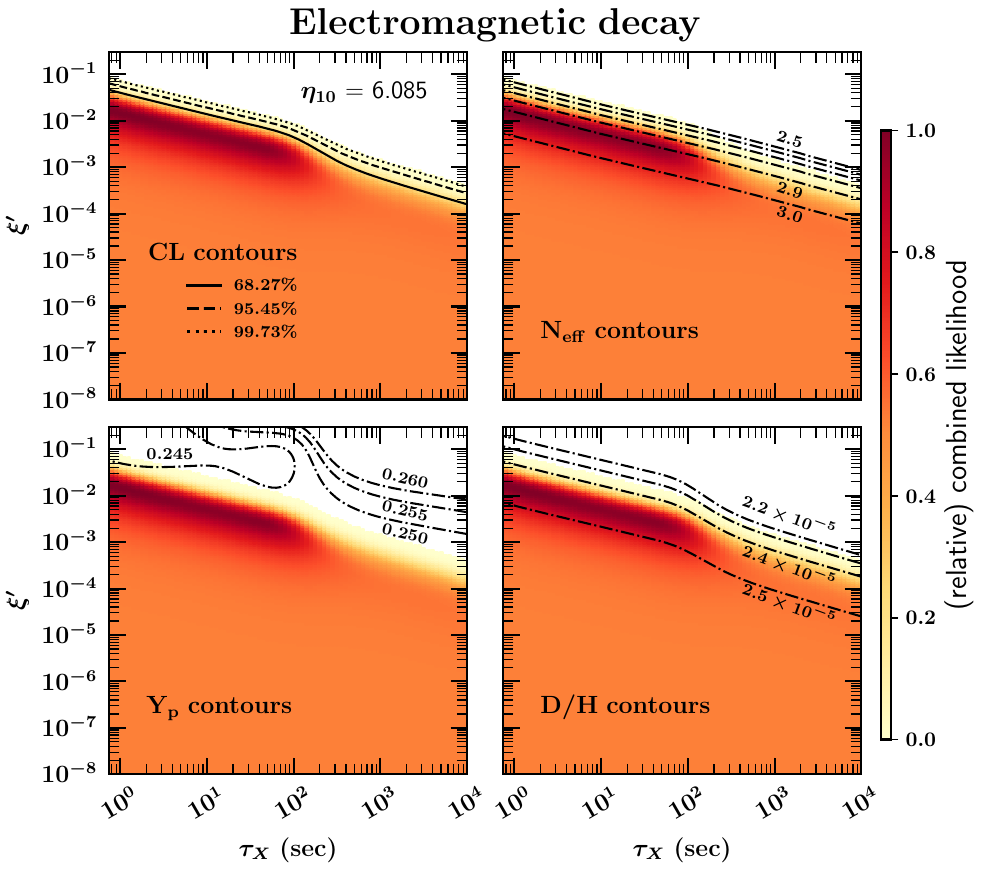}
        \caption{Contours for the EM decay case in the $(\tau_X,\xi^\prime)$ plane, evaluated at the maxim likelihood baryon-to-photon ratio $\eta_{10} = 6.085$.  Dashes lines show:  (a) likelihood contours, (b) $N_{\rm eff}$, (c) $Y_p$, (d) D/H. }
        \label{fig:EM2x2atmaxlike}
    \end{figure}

To examine the likelihood function in more detail,
in Fig.~\ref{fig:EMsliced} we show a series of likelihood plots projected onto the the $\tau_X, \xi'$ plane for different values of $\eta_{10} = 5.885-6.285$ in increments of 0.05.  Each of these corresponds to a vertical slice in Fig.~\ref{fig:EM3D}.  The peak of the likelihood function occurs at $\eta_{10} = 6.085$ (our data resolution was run at increments of 0.005 in $\eta_{10}$, so this value is essentially equivalent to the value of $\eta_{10} = 6.088$ at the peak of the SBBN likelihood function) which is shown separately in Fig.~\ref{fig:EM2x2atmaxlike}.
Since $N_\nu$ is effectively reduced (as is preferred in SBBN with variable $N_\nu$), it is not a surprise that the peak value of $\eta_{10}$ matches the SBBN with variable $N_\nu$ value. 
The total likelihood function was integrated over the $\eta, \tau_X, \xi'$ volume. Note that to properly integrate over the volume of Fig.~\ref{fig:EM3D}, the likelihood function was weighted by the product of $\tau_X \xi'$ with respect to the volume element in the linear-log-log space as
\beq
{\cal L}\ dV = {\cal L}(\eta,\tau_X, \xi')\ d\eta\ d\tau_X\ d\xi' = {\cal L} (\eta,\tau_X, \xi') \cdot \tau_X \xi' \  d\eta \ d{\rm ln} \tau_X \ d{\rm ln} \xi' \, .
\eeq
This allows us to produce the iso-likelihood contours shown in the figures. 

At values of $\eta_{10} \le 5.85$ the likelihood function is near 0 for all the values of $\tau_X, \xi'$ considered and all are outside the 99\% CL.
Starting with the upper left panel ($\eta_{10} = 5.885$) of Fig.~\ref{fig:EMsliced}, we see evidence for a non-zero likelihood along a strip of points with $\tau_X \lesssim 100$ s 
though these still lie outside the 99\% CL. 
At larger $\eta_{10} = 5.935$, the strip is expanded and is
within the 99\% CL contour shown by the dotted curve. A very thin strip is within the 95\% CL shown by the dashed curve. Moving to larger $\eta_{10} = 5.985$, we see the non-zero likelihood extending to larger $\tau_X$ and a broadened 95\% CL contour. At $\eta_{10} = 6.035$, the 68\% CL contour is visible and all points with low $\xi'$
have non-zero likelihoods.  At this and higher values of $\eta_{10}$, we see that the 95\% and 99\% CLs are no longer closed and include the Standard Model case with $\xi' = 0$ and the position of the peak of the likelihood is becoming apparent. 
At $\eta_{10} = 6.07$ (not shown) even the 68\% CL contour does not close.
The peak of the likelihood function is seen in the middle panel and in Fig.~\ref{fig:EM2x2atmaxlike}, and occurs at $\eta_{10} = 6.085, \tau_X = 0.89$~s and $\xi' = 0.016$ and the value of the likelihood function at that point is normalized to 1.  Note that at the peak, $\xi^\prime (\tau_X/1 \ \rm sec)^{1/2} = 0.015$ and the peak is very close to the left edge of the range plotted, which is $\tau_X = 0.74 - 10^4$~s where the lower limit on $\tau_X$ corresponds to decay temperatures in excess of 1 MeV. 
As one can see the fit is almost equally as good for any $\tau_X \lesssim 100$~s with $\xi'$ scaling as given in Eq.~(\ref{rest}). It is important to note that at 68\% CL, we have only an upper limit on $\xi'$ (shown by the solid curve in Fig.~\ref{fig:EMsliced}) as $\xi' = 0$ (i.e. standard BBN) is consistent with the CMB and observational data. 
When $\eta$ is increased beyond the value at the peak, as shown in the remaining panels of Fig.~\ref{fig:EMsliced}, the likelihood drops. For $\eta_{10} = 6.185$ only the 95\% and 99\% CLs are visible (even at low values of $\xi'$) and both of these likelihood contours are gone at $\eta_{10} = 6.285$.

In Fig.~\ref{fig:EM2x2atmaxlike}, we concentrate on the peak likelihood value in the $\tau_X,\xi'$ plane along the slice at $\eta_{10} = 6.085$. In addition to the likelihood contours (upper left), we show the the effective number of relativistic degrees of freedom (upper right) and the abundances of $Y_p$ (lower left) and D/H (lower right). As discussed previously, and exhibited analytically in Eqs.~(\ref{a1}-\ref{a3}), the effective of degrees of neutrino  drops below 3, as $\xi'$ is increased (for $\xi' = 0$, we have the Standard Model value of $N_{\rm eff} = 3.044$. We see then that the ridge including the peak likelihood aligns with $N_{\rm eff} \simeq 2.904$ and is close to the Standard BBN best fit value of $N_\nu = 2.898$, corresponding to $N_{\rm eff} = 2.941$; the two are slightly different 
because the matter component changes the expansion rate somewhat differently from radiation.
As one can see, the peak of the likelihood function agrees quite well with the rough estimate given in Eq.~(\ref{rest}).
While one may be tempted to conclude that BBN prefers some amount of non-relativistic (unstable) matter present during nucleosynthesis, the statistical significance of this conclusion is rather weak. The abundances of \he4 and D/H are dependent on $\eta$, but predominantly in a similar manner as in standard BBN.  We have verified that \li7 (not shown) takes standard BBN values in the allowed regions.

As noted earlier, if we fix the final value of $\eta$, then because of the dilution from $X$ decays and given in Eq.~(\ref{eq:etachange}), the value of $\eta$ during BBN may be much higher, particularly for large $\xi'$. This accounts for the increase (decrease) in the \he4 (D/H) abundance. 
For the case of $Y_p$, we see that the abundance dips to a valley for $\tau_X \lesssim 100~\rm sec$ and $\xi^\prime \sim 0.1-1$.  This corresponds to decays during BBN, when the $X$ density begins to approach the radiation density.
The \he4 abundance is sensitive to both the higher initial $\eta$ value, but also smaller $N_{\rm eff}$ value, effects which oppose each other.  In the valley region, the $N_{\rm eff}$ effect is slightly more important.

\begin{figure}[!htb]
  \centering 
\includegraphics[width=0.5\textwidth]{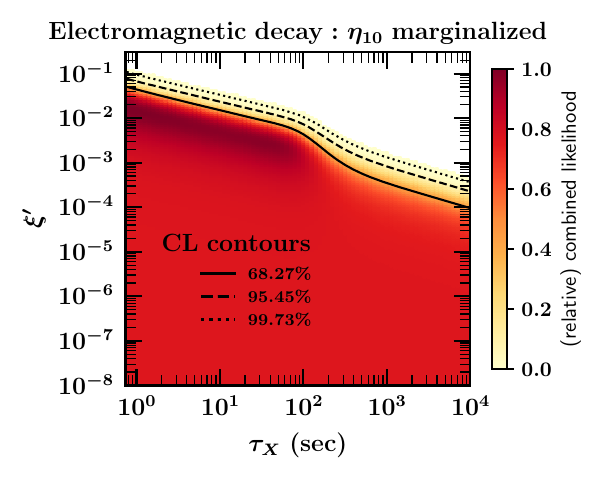}
    \caption{The likelihood function (\ref{BBNCMBobs}) for the EM case, marginalized over $\eta_{10})$  projected onto the the $(\tau_X, \xi')$ plane. 
    }
    \label{fig:Lmarg}
  \end{figure}

Figure~\ref{fig:Lmarg} show the likelihood function 
(\ref{BBNCMBobs}) marginalized over $\eta_{10}$ by integrating along the $\eta_{10}$ axis of Fig.~\ref{fig:EM3D}.
The result of marginalization over $\eta$ is projected onto the the $(\tau_X, \xi')$ plane.
As expected, we see that the contours closely follow the lines of constant $\xi^\prime \tau_X^{1/2}$ but with different amplitudes for low vs high $\tau_X$.
For $\tau_X \lesssim 600~\rm sec$ the decays happen during BBN, and we have $\xi^\prime (\tau_X/1 \rm sec)^{1/2} \lesssim 0.05$ at 68\% CL. The maximum likelihood is around $\xi^\prime (\tau_X/1 \rm sec)^{1/2} \simeq 0.015$, which is close to our estimate in eq.~(\ref{rest}).  
Overall, the marginalization looks very similar to the slices shown in Figs.~\ref{fig:EMsliced} and \ref{fig:EM2x2atmaxlike}.

\subsection{Case (b): Invisible decays}

We now turn to the case of invisible decays, where the $X$ particle
goes to neutrinos or other particles that do not interact with the plasma. 
As we have seen in \S\ref{formal}, in this case the decays do not affect the baryon-to-photon ratio and do not add to the plasma entropy, and thus do not dilute $N_\nu$ as we found in the previous section. Instead, $N_{\rm eff}$ {\em increases} in this
scenario, so we should not expect to find the improved fit for nonzero $X$ density as we did for the electromagnetic decays.

\begin{figure}[!ht]
        \centering
        \includegraphics[width=0.475\textwidth]{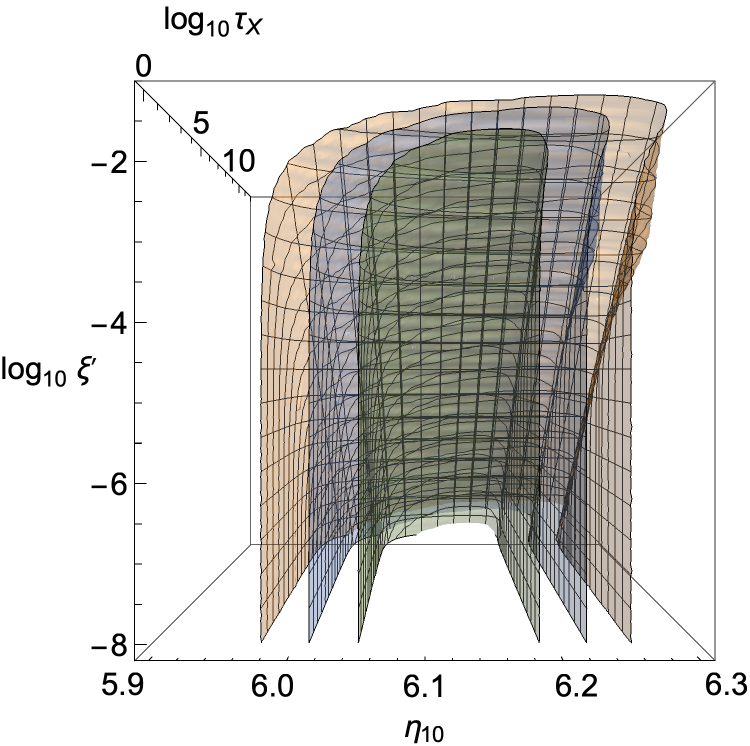}
        \includegraphics[width=0.475\textwidth]{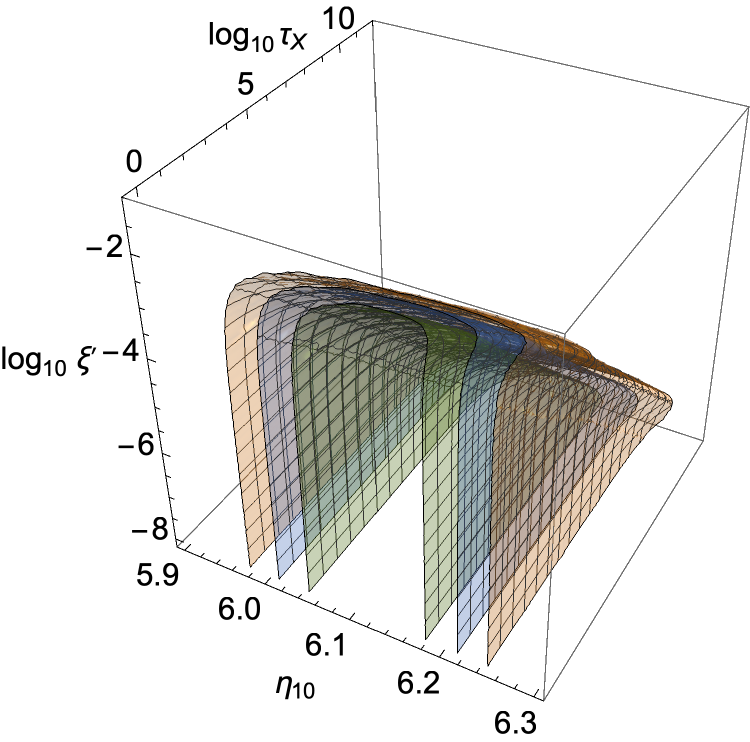}
        \caption{Likelihood contours rendered in the 3-D $(\eta,\tau_X,\xi^\prime)$ space for the case of dark decays.  We see that the CMB-preferred $\eta$ value leads to a nearly planar region of high likelihood in $(\tau_X,\xi^\prime)$.  There is a slight shift to higher $\eta$ at high $\xi^\prime$, where the likelihood is maximum. }
        \label{fig:Dark3D}
    \end{figure}

Two views of the iso-likelihood function contours for a matter component with invisible decays is shown in Fig.~\ref{fig:Dark3D}.
It bears several similarities with the corresponding figure for EM decays in that the likelihood contours become nearly vertical walls at low $\xi'$ and the relation between $\xi'$ and $\tau_X$ is maintained. At large $\xi'$, however the tilt towards lower $\eta_{10}$ at large $\xi'$ is absent and even tilts slightly toward higher $\eta_{10}$ to compensate for the change in $N_{\rm eff}$. This is due again to the positive correlation between $\eta$ and $N_\nu$ in the CMB data. 
The higher value of $N_\nu$ requires are (slightly) higher value of $\eta_{10}$. 

\begin{figure}[!ht]
        \centering
        \includegraphics[width=0.95\textwidth]{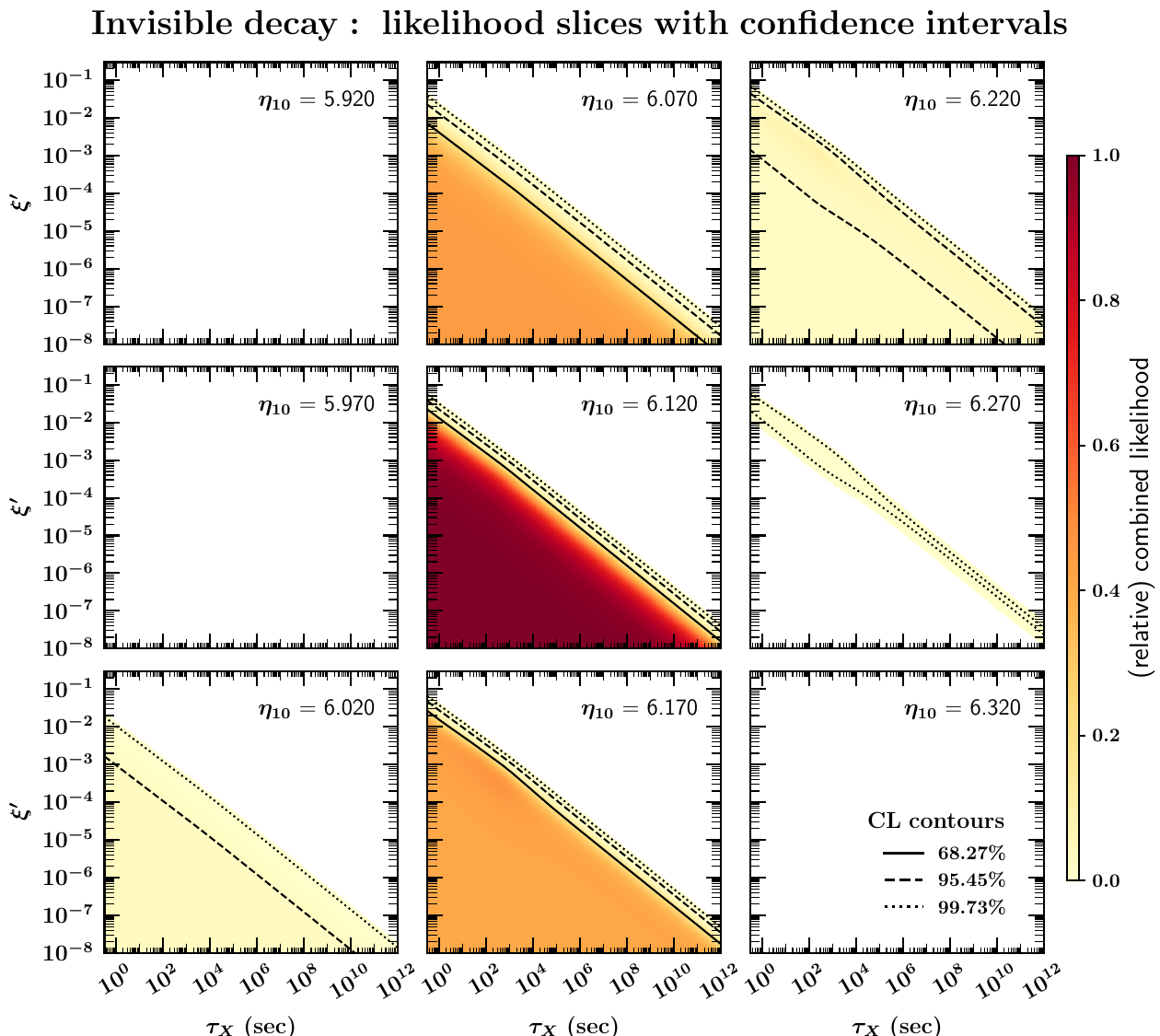}
        \caption{Likeihood contours in the $(\eta,\xi^\prime,\tau_X)$ space, shown as slices at fixed $\eta$ for the case of dark decays. Colors denote the relative value of the likelihood function indicated by the scale at the right. The 68\%, 95\% and 99\% CL contours are shown by solid, dashed, and dotted curves respectively.   }
        \label{fig:Darksliced}
    \end{figure}

Slices of this likelihood function at fixed $\eta$ are shown in Fig.~\ref{fig:Darksliced}.
In this case the peak of the likelihood function occurs at $\eta_{10} = 6.12$, i.e., slightly higher than in the EM-decay case, again due to the correlation between $\eta$ and $N_\nu$ in the CMB data. Because these decays increase $N_\nu$, the preferred value of $\xi'$ is 0. In this case the final value of $\eta$ resembles that of standard BBN where $\eta_{10} = 6.115 \pm 0.038$ \cite{ysof}.  The allowed range in $\eta_{10}$ is rather narrow and the at $\eta_{10} = 5.97$ no part of the $\tau_X, \xi'$ plane is acceptable at the 99\% CL.
Since our results are very insensitive to $\xi'$ and $\tau_X$
when $\xi' \ll 1$, we choose to normalize the likelihood function $\mathcal{L}=1$ at $\xi' = 10^{-8}$ and $\tau_X = 1$~s.
At $\eta_{10} = 6.02$, we see both 99\% and 95\% contours and the 68\% CL contour appears at $\eta_{10} = 6.07$. At $\eta_{10} = 6.32$
again, no part of the plane is acceptable at even 99\% CL. Note that because the decays are invisible in this case, they can not affect the light element abundances once BBN is complete. This is contrast to the case of EM decays where decays with lifetimes longer than $\sim 10^4$~s can affect the light element abundances and a different analysis is needed. 

\begin{figure}[!ht]
        \centering
        \includegraphics[width=0.8\textwidth]{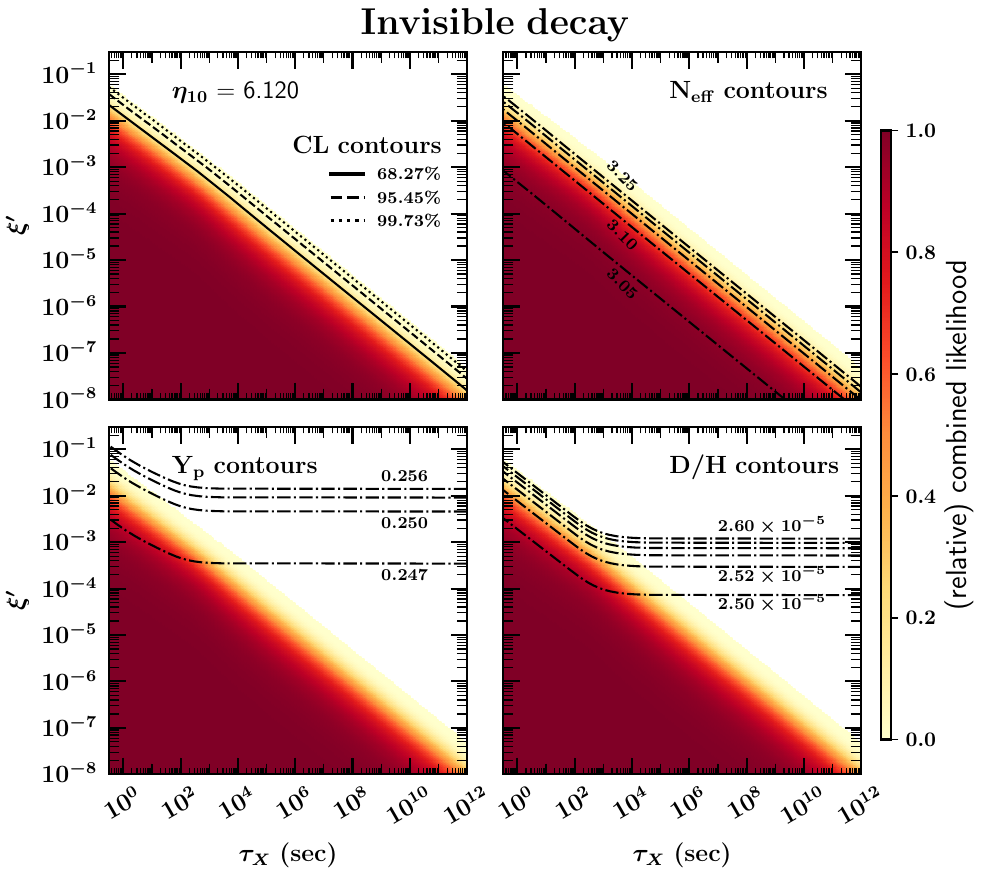}
        \caption{Contours for the case of invisible decays in the $(\tau_X,\xi^\prime)$ plane, evaluated at the maxim likelihood baryon-to-photon ratio $\eta_{10} = 6.120$.  Dashes lines show:  (a) likelihood contours, (b) $N_{\rm eff}$, (c) $Y_p$, (d) D/H. 
        }
        \label{fig:Dark2x2atmaxlike}
    \end{figure}

\begin{figure}[!htb]
  \centering 
\includegraphics[width=0.5\textwidth]{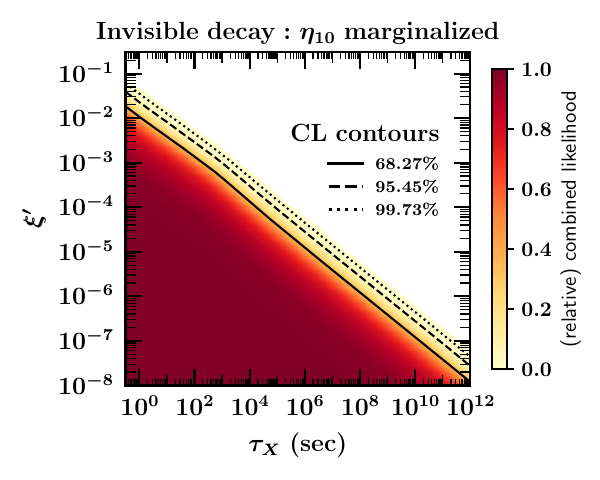}
    \caption{The likelihood function (\ref{BBNCMBobs}) for the invisible decays  marginalized over $\eta_{10})$ and projected onto the the $(\tau_X, \xi')$ plane. 
    }
    \label{fig:DarkLmarg}
  \end{figure}

The slice at the peak value of $\eta_{10} = 6.12$ showing contours of the $N_{\rm eff}$ and the abundances of \he4 and D/H are shown in Fig.~\ref{fig:Dark2x2atmaxlike}. As expected,
we see the increase in $N_{\rm eff}$ as $\xi'$ is increased.  
For $Y_p$ we see that at small $\tau_X$ and high $\xi^\prime$, the constraint roughly follows a similar slope as the $N_{\rm eff}$ contour.
But for $\tau_X \gtrsim 100$ sec, the contour becomes horizontal, independent of $\xi^\prime$.  In this regime, the decays occur after BBN, and beneath the $Y_p$ contour, the perturbations are small during BBN.  Thus, nucleosynthesis proceeds as in the standard case with $N_\nu = 3$, and the light elements give the usual results.
However, the decays still affect the CMB $N_{\rm eff}$, which dominate the constraints in this regime. The abundance of D/H, exhibits similar behavior as $Y_p$. Finally, we show in Fig.~\ref{fig:DarkLmarg},
    the likelihood plot projecton onto the $\tau_X, \xi'$ plane after marginalizing over $\eta_{10}$.   

\section{Discussion}
\label{summary}

The excellent concordance between BBN theory, the observational determination of the \he4 and D/H abundances, and observations of the CMB anisotropy within the context of the standard models of particle and nuclear physics and cosmology enable us to set strong constraints on any departures from Standard Model physics. A common example of the strength of this concordance is its ability to 
constrain the effective number of relativistic degrees of freedom or the number of neutrino flavors. As recently shown in \cite{ysof},
when the number of neutrinos is allowed to deviate from its standard model value of 3, 
the peak BBN-CMB convolved likelihood result is $N_\nu = 2.898 \pm 0.141$.

Here we look at another relatively simple and well-motivated scenario that perturbs BBN: the presence of a species $X$ that acts as matter and then decays out of equilibrium during or after nucleosynthesis.  The presence of a matter component changes the expansion history in ways not captured by the addition of relativistic species, but if the perturbations are small we find that the net change to
$N_{\rm eff}$ provides rather accurate insight into the constraints posed by the light elements and the CMB.

We consider both electromagnetic as well as dark decays, which must be treated separately.
Interestingly, the electromagnetic or visible decays lead to a dilution of 
the neutrino energy density, and thus a decrease in $N_{\rm eff}$.
In addition, the decays reduce $\eta$, so that its initial value must be higher than usual in order to evolve to the CMB-preferred range.  These two features
combine with the mild CMB preference for $N_{\rm eff} < 3$ to lead to a locus of nonzero perturbations giving the best fit.  
This regime is well-described by $\xi^\prime (\tau_X/1 \, \rm sec)^{1/2} \sim$ 0.015, and $\tau_X \sim 1$ to 100 sec.
To be sure, the statistical preference for this case is mild, and the standard BBN case still provides an excellent fit, with large regions of parameter space ruled out.  

For the case of dark/invisible decays, the evolution of $\eta$  is not perturbed, and the main effects can largely be understood in terms of changing $N_{\rm eff}$.
For lifetimes $\tau_X \lesssim 100$ sec, the decays occur during BBN, and the light elements place constraints competitive with the CMB $N_{\rm eff}$.  At larger lifetimes the light elements are unaffected during BBN, and then the CMB $N_{\rm eff}$ constraints dominate the limits.

We look forward to future measurements that will tighten these limits.
CMB-S4 should substantially improve the precision of the CMB-determined $N_{\rm eff}$ \cite{CMB-S4}, which as we have seen plays an important role in all of our constraints and a dominant role for EM constraints at $\tau_X \gtrsim 100$ sec.  Ongoing campaigns to observe \he4 in low-metallicity dwarf galaxies can improve $Y_p$. 
Progress in this direction began with \cite{abopss,Aver:2021rwi} and is ongoing. And finally, nuclear physics measurements of the $d+d$ cross section can reduced the theoretical D/H uncertainties that currently dominate the deuterium error budget \cite{dpg,ysof}, making it a more powerful probe of new physics.

\acknowledgments

TRIUMF receives federal funding via a contribution agreement with the National Research Council of Canada. The work of K.A.O.~was supported in part by DOE grant DE-SC0011842 at the University of Minnesota.

\end{document}